# Singular Layer Transmission for Continuous-Variable Quantum Key Distribution

Laszlo Gyongyosi

[1] Quantum Technologies Laboratory, Department of Telecommunications
*Budapest University of Technology and Economics*
2 Magyar tudosok krt, Budapest, *H*-1117, Hungary
[2] Information Systems Research Group, Mathematics and Natural Sciences
*Hungarian Academy of Sciences*
Budapest, *H*-1518, Hungary

gyongyosi@hit.bme.hu

**Abstract**

We develop a singular layer transmission model for continuous-variable quantum key distribution (CVQKD). In CVQKD, the transmit information is carried by continuous-variable (CV) quantum states, particularly by Gaussian random distributed position and momentum quadratures. The reliable transmission of the quadrature components over a noisy link is a cornerstone of CVQKD protocols. The proposed singular layer uses the singular value decomposition of the Gaussian quantum channel, which yields an additional degree of freedom for the phase space transmission. This additional degree of freedom can further be exploited in a multiple-access scenario. The singular layer defines the eigenchannels of the Gaussian physical link, which can be used for the simultaneous reliable transmission of multiple user data streams. Our transmission model also includes the singular interference avoider (SIA) precoding scheme. The proposed SIA precoding scheme prevents the eigenchannel interference to reach an optimal transmission over a Gaussian link. We demonstrate the results through the adaptive multicarrier quadrature division–multiuser quadrature allocation (AMQD-MQA) CVQKD multiple-access scheme. We define the singular model of AMQD-MQA and characterize the properties of the eigenchannel interference. We propose the SIA precoding of Gaussian random quadratures and the optimal decoding at the receiver. We show a random phase space constellation scheme for the Gaussian sub-channels. The singular layer transmission provides improved simultaneous transmission rates for the users with unconditional security in a multiple-access scenario, particularly in crucial low signal-to-noise ratio regimes.

# 1 Introduction

The continuous-variable quantum key distribution (CVQKD) protocols allow for the legal parties to transmit information with unconditional security over the current, well-established standard telecommunication networks. The CVQKD protocols use continuous quantum systems for the information transmission, practically Gaussian random distributed position and momentum quadratures in the phase space [1-15]. These quadratures identify a continuous-variable (CV) quantum state in the phase space, which are then transmitted over a noisy quantum link (e.g., via an optical fiber or a wireless optical channel [16-17]), which is attacked by an eavesdropper [13-14]. Because the optimal attack against CVQKD protocols is a Gaussian attack, the noise of the physical quantum link can be provably modeled as an additive white Gaussian noise and the link as a Gaussian channel. The CVQKD protocols are equipped with several benefits in comparison with the discrete-variable (DV) QKD; however, in comparison with the traditional telecommunication protocols, the efficiency of CVQKD still requires significant improvements. An enhancement is particularly crucial and essential for the low signal-to-noise ratio (SNR) regimes, at which the experimental long-distance CVQKD protocols are operating. For this purpose, the adaptive multicarrier quadrature division (AMQD) modulation scheme has been recently introduced for CVQKD [4], which injects several convenient abilities into CVQKD regarding the transmission rates, distances, and tolerable noise level, similar to the orthogonal frequency-division multiplexing (OFDM) [10] of traditional networking. In particular, the AMQD modulation granulates the transmit information into Gaussian subcarrier CVs and formulates Gaussian sub-channels from the physical Gaussian link, leading to improved transmission distances, higher secret key rates, and enhanced tolerable excess noise. The benefits of AMQD have also been extended to a multiuser scenario within the framework of the AMQD–multiuser quadrature allocation (MQA) multiple-access scheme. The AMQD-MQA [3] uses the Gaussian subcarriers and the AMQD modulation for the reliable simultaneous transmission of the users' input CVs and provides capacity-achieving multiple-access communication for the parties over a shared physical Gaussian quantum channel.

The singular value decomposition (SVD) is a well-known tool in linear algebra, with a widespread application from mathematical statistic to signal processing. In this work, we show that the benefits of SVD can also be exploited in a CVQKD scenario. We reveal that the information transmission of CVQKD over the Gaussian quantum channel can be rephrased in terms of SVD, which is called the *singular layer* transmission model throughout. The singular layer injects an additional degree of freedom into the transmission, which can be exploited in the protocol. We define the singular layer of AMQD and AMQD-MQA and prove that the singular layer allows for the parties to reach improved transmission distances and higher secret key rates. Specifically, the singular layer defines a higher-level layer above the physical layer and transforms the quantum channel onto a set of *eigenchannels*. From the layer structure, it follows that the singular layer transmission does not require any change in the lower layers, that is, the physical layer transmission remains untouched, because the SVD defines a purely logical layer over the physical layer transmission. In Section 3, we reveal how the singular layer transmission can be exploited in the multicarrier-level transmission of AMQD. We also define the singular layer model of AMQD-MQA multiple access, and by exploiting the tools of the quantum Shannon theory, we study the achievable reliable transmission rates. We characterize the optimal phase space decoding process and an information-theoretical optimal decoder for the Gaussian random phase space quadratures. At a partial channel side information (e.g., an exact state information about the Gaussian sub-channels is not available for the parties), the eigenchannels can *interfere* with one another; thus, the handling of an interference avoider is a crucial point in the SVD-level approach of a CVQKD protocol. Because the interference cancellation, in theory, could require an unlimited modulation variance because the interference vector could occur in the arbitrary domains of the phase space, a carefully designed interference avoider is a critical requirement on the SVD-level. For this purpose, in Section 4, we define the *singular interference avoider* (SIA) scheme. In Section 5, we show a random phase space coding scheme for the transmission of Gaussian subcarrier CVs over the Gaussian sub-channels, which provides an unconditional security and an optimal partition of the phase space. The benefits of the proposed singular layer model can be exploited most significantly in the crucial low-SNR regimes, which is in fact particularly important in the implementation of long-distance CVQKD.

This paper is organized as follows. Section 2 summarizes the preliminary findings. Section 3 discusses the singular layer transmission model. Section 4 studies the SIA scheme and the optimal phase space decoding process. Section 5 defines a phase space coding scheme for the transmission of Gaussian subcarrier CVs. Finally, Section 6 concludes the results. Supplemental information is included in the Appendix.

## 2 Preliminaries

### 2.1 Basic terms and notations

Before we would start to discuss the singular layer model, we summarize the basic notations of the AMQD-MQA multiple access CVQKD scheme from Section 2.1 of [3]. Further related background material can be found in [4.]

The input of $k$-th user $U_k$ is a Gaussian CV state $|\psi\rangle_k \in \mathcal{S}$, where $\mathcal{S}$ stands for the phase space. The noise of the Gaussian quantum channel $\mathcal{N}$ is denoted by $\Delta$, which acts independently on the $x$ position, and $p$ momentum quadratures. These Gaussian CV states, along with the noise of the quantum channel, can be modeled as Gaussian random continuous-variables.

A *single-carrier* Gaussian modulated CV state $|\psi\rangle$ in the phase space $\mathcal{S}$ is modeled as a zero-mean, circular symmetric complex Gaussian random variable $z \in \mathcal{CN}(0, \sigma_z^2)$, with i.i.d. zero mean, Gaussian random quadrature components $x, p \in \mathbb{N}(0, \sigma_{\omega_0}^2)$, where $\sigma_{\omega_0}^2$ is the modulation variance. The variance of $z$ is

$$\sigma_z^2 = \mathbb{E}\left[|z|^2\right] = 2\sigma_{\omega_0}^2. \tag{1}$$

The $\Delta \in \mathcal{CN}(0,\sigma_\Delta^2)$ noise variable of the Gaussian channel $\mathcal{N}$ with i.i.d. zero-mean, Gaussian random components on the position and momentum quadratures $\Delta_x, \Delta_p \in \mathbb{N}(0,\sigma_\mathcal{N}^2)$ has the variance of

$$\sigma_\Delta^2 = \mathbb{E}\left[|\Delta|^2\right] = 2\sigma_\mathcal{N}^2. \quad (2)$$

A $|\phi\rangle \in \mathcal{S}$ Gaussian *subcarrier* CV state is also modeled by a zero-mean, circular symmetric Gaussian random variable $d \in \mathcal{CN}(0,\sigma_d^2)$, with i.i.d. zero mean, Gaussian random quadrature components $x_d, p_d \in \mathbb{N}(0,\sigma_\omega^2)$, where $\sigma_\omega^2$ is the modulation variance of the Gaussian subcarrier CV state. The Gaussian subcarrier random variable $d$ has variance

$$\sigma_d^2 = \mathbb{E}\left[|d|^2\right] = 2\sigma_\omega^2. \quad (3)$$

Let assume that $K$ independent users are in the CVQKD system.

The subset of *allocated* users is denoted by $\mathcal{A} \subseteq K$. Only the allocated users can transmit information in a given (particularly $j$-th) AMQD block. An AMQD block consist of $l$ Gaussian subcarrier CVs (assuming an optimal Gaussian collective attack [13-13], only $l$ sub-channels have high quality from the $n$, these are referred as *good* sub-channels throughout, for details see [3] and [4]). Following the formalism of AMQD [4], the variables of an allocated user $U_k$, $k = 1,\ldots,|\mathcal{A}|$, where $|\mathcal{A}|$ is the cardinality of the subset $\mathcal{A}$, are as follows. The $i$-th Gaussian modulated input coherent state of $U_k$ is referred as $|\varphi_{k,i}\rangle = |x_{k,i} + \mathrm{i}p_{k,i}\rangle$, where $x_{k,i} \in \mathbb{N}(0,\sigma_{\omega_{0,k}}^2)$, $p_{k,i} \in \mathbb{N}(0,\sigma_{\omega_{0,k}}^2)$ are the position and momentum quadratures with variance $\sigma_{\omega_{0,k}}^2$, respectively. This CV state can be rewritten as a zero-mean, circular symmetric complex Gaussian random variable $z_{k,i} \in \mathcal{CN}(0,\sigma_{\omega_{z_{k,i}}}^2)$, $\sigma_{\omega_{z_{k,i}}}^2 = \left[\mathbb{E}|z_{k,i}|^2\right]$, as

$$z_{k,i} = x_{k,i} + \mathrm{i}p_{k,i}, \quad (4)$$

thus

$$|\varphi_{k,i}\rangle = |z_{k,i}\rangle. \quad (5)$$

The variable $e^{\mathrm{i}\varphi_i}z_{k,i}$ has the same distribution of $z_{k,i}$ for any $\varphi_i$, i.e., $\mathbb{E}[z_{k,i}] = \mathbb{E}[e^{\mathrm{i}\varphi_i}z_{k,i}] = \mathbb{E}e^{\mathrm{i}\varphi_i}[z_{k,i}]$ and $\sigma_{z_{k,i}}^2 = \mathbb{E}\left[|z_{k,i}|^2\right]$. The density of $z_{k,i}$ is

$$f(z_{k,i}) = \frac{1}{2\pi\sigma_{\omega_{0,k}}^2} e^{\frac{-\left(|z_{k,i}|^2\right)}{2\sigma_{\omega_{0,k}}^2}} = f(x_{k,i}, p_{k,i}) = \frac{1}{2\pi\sigma_{\omega_{0,k}}^2} e^{\frac{-\left(x_{k,i}^2 + p_{k,i}^2\right)}{2\sigma_{\omega_{0,k}}^2}}, \quad (6)$$

where $|z_{k,i}| = \sqrt{x_{k,i}^2 + p_{k,i}^2}$ is the magnitude, which is a Rayleigh random variable with density

$$f\left(\left|z_{k,i}\right|\right) = \frac{\left|z_{k,i}\right|}{\sigma^2_{\omega_{z_{k,i}}}} e^{\frac{-\left|z_{k,i}\right|^2}{2\sigma^2_{\omega_{z_{k,i}}}}}, \left|z_{k,i}\right| \geq 0, \tag{7}$$

while the $\left|z_{k,i}\right|^2 = x^2_{k,i} + p^2_{k,i}$ squared magnitude is exponentially distributed with density

$$f\left(\left|z_{k,i}\right|^2\right) = \frac{1}{\sigma^2_{\omega_{z_{k,i}}}} e^{\frac{-\left|z_{k,i}\right|^2}{\sigma^2_{\omega_{z_{k,i}}}}}, \left|z_{k,i}\right|^2 \geq 0. \tag{8}$$

The $i$-th Gaussian *subcarrier* CV of user $U_k$ is defined as

$$\left|\phi_i\right\rangle = \left|\text{IFFT}\left(z_{k,i}\right)\right\rangle = \left|F^{-1}\left(z_{k,i}\right)\right\rangle = \left|d_i\right\rangle, \tag{9}$$

where IFFT stands for the Inverse Fast Fourier Transform, and subcarrier continuous-variable $\left|\phi_i\right\rangle$ in Equation (9) is also a zero-mean, circular symmetric complex Gaussian random variable $d_i \in \mathcal{CN}\left(0, \sigma^2_{d_i}\right)$, $\sigma^2_{d_i} = \mathbb{E}\left[\left|d_i\right|^2\right]$. The quadrature components of the modulated Gaussian subcarrier CVs are referred by $x_d, p_d \in \mathbb{N}\left(0, \sigma^2_\omega\right)$, where $\sigma^2_\omega$ is the constant modulation variance of AMQD that is used in the transmission phase (*Note*: the *constant* modulation variance is provably the optimal solution in the low-SNR regimes, because the performance is very close to the exact allocation [3-4], [18-19], [21-23]. This property will not change in the singular layer of AMQD and AMQD-MQA.).

Assuming $K$ independent users in the AMQD-MQA, who transmit the single-carrier Gaussian CVs to an encoder $\mathcal{E}$, (9) modifies as follows

$$\left|\phi_i\right\rangle = \left|\text{CVQFT}^\dagger\left(z_{k,i}\right)\right\rangle = \left|F^{-1}\left(z_{k,i}\right)\right\rangle = \left|d_i\right\rangle, \tag{10}$$

where CVQFT$^\dagger$ refers to the continuous-variable inverse quantum Fourier transform (QFT).

The inverse of (9) results the single-carrier CV from the subcarrier CV as follows:

$$\left|\varphi_{k,i}\right\rangle = \text{CVQFT}\left(\left|\phi_i\right\rangle\right) = F\left(\left|d_i\right\rangle\right) = \left|F\left(F^{-1}\left(z_{k,i}\right)\right)\right\rangle = \left|z_{k,i}\right\rangle, \tag{11}$$

where CVQFT is the continuous-variable QFT operation.

Let $\mathbf{z}_k$ be a $d$-dimensional, zero-mean, circular symmetric complex random Gaussian vector of $U_k$,

$$\mathbf{z}_k = \mathbf{x}_k + \mathrm{i}\mathbf{p}_k = \left(z_{k,1}, \ldots, z_{k,d}\right)^T \in \mathcal{CN}\left(0, \mathbf{K}_{\mathbf{z}_k}\right), \tag{12}$$

where $\mathbf{K}_{\mathbf{z}_k}$ is the $d \times d$ Hermitian covariance matrix of $\mathbf{z}_k$, $\mathbf{K}_{\mathbf{z}_k} = \mathbb{E}\left[\mathbf{z}_k \mathbf{z}_k^\dagger\right]$, and $\mathbf{z}_k^\dagger$ stands for the adjoint of $\mathbf{z}_k$. Each $z_{k,i}$ variable is a zero-mean, circular symmetric complex Gaussian random variable $z_{k,i} \in \mathcal{CN}\left(0, \sigma^2_{\omega_{z_{k,i}}}\right)$, $z_{k,i} = x_{k,i} + \mathrm{i}p_{k,i}$. The real and imaginary variables (i.e., the position and momentum quadratures) formulate *d*-dimensional real Gaussian random vectors, $\mathbf{x}_k = (x_{k,1},...,x_{k,d})^T$ and $\mathbf{p}_k = (p_{k,1},...,p_{k,d})^T$, with zero-mean Gaussian random variables

$$x_{k,i} = \frac{1}{\sigma_{\omega_{0,k}}\sqrt{2\pi}} e^{\frac{-x_{k,i}^2}{2\sigma_{\omega_{0,k}}^2}}, \; p_{k,i} = \frac{1}{\sigma_{\omega_{0,k}}\sqrt{2\pi}} e^{\frac{-p_{k,i}^2}{2\sigma_{\omega_{0,k}}^2}}, \quad (13)$$

where $\sigma_{\omega_{0,k}}^2$ is the stands for single-carrier modulation variance (precisely, the variance of the real and imaginary components of $z_{k,i}$). For vector $\mathbf{z}_k$,

$$\mathbb{E}[\mathbf{z}_k] = \mathbb{E}[e^{i\gamma}\mathbf{z}_k] = \mathbb{E}e^{i\gamma}[\mathbf{z}_k] \quad (14)$$

holds, and

$$\mathbb{E}[\mathbf{z}_k \mathbf{z}_k^T] = \mathbb{E}[e^{i\gamma}\mathbf{z}_k (e^{i\gamma}\mathbf{z}_k)^T] = \mathbb{E}e^{i2\gamma}[\mathbf{z}_k \mathbf{z}_k^T], \quad (15)$$

for any $\gamma \in [0, 2\pi]$. The density of $\mathbf{z}_k$ is as follows (if $\mathbf{K}_{\mathbf{z}_k}$ is invertible):

$$f(\mathbf{z}_k) = \frac{1}{\pi^d \det \mathbf{K}_{\mathbf{z}_k}} e^{-\mathbf{z}_k^\dagger \mathbf{K}_{\mathbf{z}_k}^{-1} \mathbf{z}_k}. \quad (16)$$

A *d*-dimensional Gaussian random vector is expressed as $\mathbf{x}_k = \mathbf{As}$, where $\mathbf{A}$ is an (invertible) linear transform from $\mathbb{R}^d$ to $\mathbb{R}^d$, and $\mathbf{s}$ is a *d*-dimensional standard Gaussian random vector $\mathbb{N}(0,1)_d$. This vector is characterized by its covariance matrix $\mathbf{K}_{\mathbf{x}_k} = \mathbb{E}[\mathbf{x}_k \mathbf{x}_k^T] = \mathbf{AA}^T$, as

$$\mathbf{x}_k = \frac{1}{(\sqrt{2\pi})^d \sqrt{\det(\mathbf{AA}^T)}} e^{-\frac{\mathbf{x}_k^T \mathbf{x}_k}{2(\mathbf{AA}^T)}}. \quad (17)$$

The Fourier transformation $F(\cdot)$ of an *l*-dimensional Gaussian random vector $\mathbf{v} = (v_1,...,v_l)^T$ results in the *d*-dimensional Gaussian random vector $\mathbf{m} = (m_1,...,m_d)^T$:

$$\mathbf{m} = F(\mathbf{v}) = e^{\frac{-\mathbf{m}^T \mathbf{AA}^T \mathbf{m}}{2}} = e^{\frac{-\sigma_{\omega_0}^2 (m_1^2+...+m_d^2)}{2}}. \quad (18)$$

In the first step of AMQD, Alice applies the inverse FFT operation to vector $\mathbf{z}_k$ (see Equation (12)), which outputs an *l*-dimensional zero-mean, circular symmetric complex Gaussian random vector $\mathbf{d}$, $\mathbf{d} \in \mathcal{CN}(0, \mathbf{K_d})$, $\mathbf{d} = (d_1,...,d_l)^T$, as

$$\mathbf{d} = F^{-1}(\mathbf{z}_k) = e^{\frac{\mathbf{d}^T \mathbf{AA}^T \mathbf{d}}{2}} = e^{\frac{\sigma_{\omega_0}^2 (d_1^2+...+d_l^2)}{2}}, \quad (19)$$

where $d_i = x_{d_i} + ip_{d_i}$, $d_i \in \mathcal{CN}(0, \sigma_{d_i}^2)$, and the position and momentum quadratures of $|\phi_i\rangle$ are i.i.d. Gaussian random variables

$$x_{d_i} \in \mathbb{N}\left(0, \sigma_F^2\right), \ p_{d_i} \in \mathbb{N}\left(0, \sigma_F^2\right), \tag{20}$$

where $\mathbf{K_d} = \mathbb{E}\left[\mathbf{dd}^\dagger\right]$, $\mathbb{E}[\mathbf{d}] = \mathbb{E}\left[e^{\mathrm{i}\gamma}\mathbf{d}\right] = \mathbb{E}e^{\mathrm{i}\gamma}[\mathbf{d}]$, and $\mathbb{E}\left[\mathbf{dd}^T\right] = \mathbb{E}\left[e^{\mathrm{i}\gamma}\mathbf{d}\left(e^{\mathrm{i}\gamma}\mathbf{d}\right)^T\right] = \mathbb{E}e^{\mathrm{i}2\gamma}\left[\mathbf{dd}^T\right]$ for any $\gamma \in [0, 2\pi]$. The coherent Gaussian subcarrier CV is

$$\left|\phi_i\right\rangle = \left|d_i\right\rangle = \left|F^{-1}(z_k)\right\rangle. \tag{21}$$

The result of Equation (19) defines $l$ independent $\mathcal{N}_i$ Gaussian sub-channels, each with noise variance $\sigma_{\mathcal{N}_i}^2$, one for each subcarrier coherent state $\left|\phi_i\right\rangle$. After the CV subcarriers are transmitted through the noisy quantum channel, Bob applies the CVQFT, which results him the noisy version $\left|\varphi_k'\right\rangle = \left|z_k'\right\rangle$ of the input $z_k$ of $U_k$.

The $m$-th element of $d$-dimensional zero-mean, circular symmetric complex Gaussian random output vector $\mathbf{y}_k \in \mathcal{CN}\left(0, \mathbb{E}\left[\mathbf{y}_k \mathbf{y}_k^\dagger\right]\right)$ of $U_k$, is as follows:

$$\begin{aligned} y_{k,m} &= F\left(\mathbf{T}(\mathcal{N})\right)z_{k,m} + F(\Delta) \\ &= F\left(\mathbf{T}(\mathcal{N})\right)F\left(F^{-1}(z_{k,m})\right) + F(\Delta) \\ &= \sum_l F\left(T_i(\mathcal{N}_i)\right)F(d_i) + F(\Delta_i), \end{aligned} \tag{22}$$

where $F\left(\mathbf{T}(\mathcal{N})\right)$ is the Fourier transform of the $l$-dimensional complex channel transmission vector

$$\mathbf{T}(\mathcal{N}) = \left(T_1(\mathcal{N}_1)..., T_l(\mathcal{N}_l)\right)^T \in \mathcal{C}^l, \tag{23}$$

where

$$T_i(\mathcal{N}_i) = \mathrm{Re}\left(T_i(\mathcal{N}_i)\right) + \mathrm{i}\,\mathrm{Im}\left(T_i(\mathcal{N}_i)\right) \in \mathcal{C}, \tag{24}$$

is a complex variable, called *transmittance coefficient,* which quantifies the position and momentum quadrature transmission (i.e., gain) of the $i$-th Gaussian sub-channel $\mathcal{N}_i$, in the phase space $\mathcal{S}$, with (normalized) real and imaginary parts $0 \leq \mathrm{Re}\,T_i(\mathcal{N}_i) \leq 1/\sqrt{2}$, $0 \leq \mathrm{Im}\,T_i(\mathcal{N}_i) \leq 1/\sqrt{2}$. The $T_i(\mathcal{N}_i)$ variable has a magnitude of $\left|T_i(\mathcal{N}_i)\right| = \sqrt{\mathrm{Re}\,T_i(\mathcal{N}_i)^2 + \mathrm{Im}\,T_i(\mathcal{N}_i)^2} \in \mathbb{R}$, where $\mathrm{Re}\,T_i(\mathcal{N}_i) = \mathrm{Im}\,T_i(\mathcal{N}_i)$, by our convention.

For the $l$ sub-channels, the $F(\Delta)$ complex vector is evaluated as

$$F(\Delta) = e^{\frac{-F(\Delta)^T \mathbf{K}_{F(\Delta)} F(\Delta)}{2}} = e^{\frac{-\left[F(\Delta_1)^2 \sigma_{\mathcal{N}_1}^2 + ... + F(\Delta_l)^2 \sigma_{\mathcal{N}_l}^2\right]}{2}}, \tag{25}$$

which is the Fourier transform of the $l$-dimensional zero-mean, circular symmetric complex Gaussian noise vector $\Delta \in \mathcal{CN}\left(0, \sigma_\Delta^2\right)_d$, $\Delta = \left(\Delta_1, ..., \Delta_l\right)^T \in \mathcal{CN}\left(0, \mathbf{K}_\Delta\right)$, where $\mathbf{K}_\Delta = \mathbb{E}\left[\Delta\Delta^\dagger\right]$, with independent, zero-mean Gaussian random components $\Delta_{x_i} \in \mathbb{N}\left(0, \sigma^2_{\mathcal{N}_i}\right)$, $\Delta_{p_i} \in \mathbb{N}\left(0, \sigma^2_{\mathcal{N}_i}\right)$ with variance $\sigma^2_{\mathcal{N}_i}$ for each $\Delta_i$. These identify the Gaussian noise of sub-channel $\mathcal{N}_i$ on the quadrature components in the phase space $\mathcal{S}$. The CVQFT-transformed noise vector can be rewritten as $F(\Delta) = \left(F(\Delta_1), \ldots, F(\Delta_l)\right)^T$, with Fourier-transformed quadrature noise components $F(\Delta_{x_i}) \in \mathbb{N}\left(0, \sigma^2_{F(\mathcal{N}_i)}\right) = \mathbb{N}\left(0, \sigma^2_{\mathcal{N}_i}\right)$, $F(\Delta_{p_i}) \in \mathbb{N}\left(0, \sigma^2_{F(\mathcal{N}_i)}\right) = \mathbb{N}\left(0, \sigma^2_{\mathcal{N}_i}\right)$ for each $F(\Delta_i)$, which defines a $d$-dimensional zero-mean, circular symmetric complex Gaussian random vector $F(\Delta) \in \mathcal{CN}\left(0, \mathbf{K}_{F(\Delta)}\right) = \mathcal{CN}\left(0, \mathbf{K}_\Delta\right)$ with a covariance matrix

$$\mathbf{K}_{F(\Delta)} = \mathbb{E}\left[F(\Delta) F(\Delta)^\dagger\right]. \tag{26}$$

An AMQD *block* is formulated from $l$ Gaussian subcarrier continuous-variables, as follows:

$$\mathbf{y}[j] = F(\mathbf{T}(\mathcal{N})) F(\mathbf{d})[j] + F(\Delta)[j], \tag{27}$$

where $j$ is the index of the AMQD block, $F(\mathbf{d}) = F\left(F^{-1}(\mathbf{z})\right)$, for $F^{-1}(\mathbf{z})$ see (19), while

$$\begin{aligned}
\mathbf{y}[j] &= \left(y_1[j], \ldots, y_d[j]\right)^T, \\
F(\mathbf{d})[j] &= \left(F(d_1)[j], \ldots, F(d_l)[j]\right)^T, \\
F(\Delta)[j] &= \left(F(\Delta_1)[j], \ldots, F(\Delta_l)[j]\right)^T.
\end{aligned} \tag{28}$$

The squared magnitude $\tau = \|F(\mathbf{d})[j]\|^2$ is an exponentially distributed variable, with density $f(\tau) = \left(1/2\sigma^{2n}_\omega\right) e^{-\tau/2\sigma^2_\omega}$, and from the Parseval theorem [18-20] follows, that $\mathbb{E}[\tau] \leq n 2\sigma^2_\omega$, while the average quadrature modulation variance of the Gaussian subcarriers is

$$\sigma^2_\omega = \frac{1}{n}\sum_{i=1}^{n} \sigma^2_{\omega_i} = \sigma^2_{\omega_0}. \tag{29}$$

Eve's attack on sub-channel $\mathcal{N}_i$ is modeled by the $T_{Eve,i}$ normalized complex transmittance $T_{Eve,i} = \operatorname{Re} T_{Eve,i} + i \operatorname{Im} T_{Eve,i} \in \mathcal{C}$, where $0 \leq \operatorname{Re} T_{Eve,i} \leq 1/\sqrt{2}$, $0 \leq \operatorname{Im} T_{Eve,i} \leq 1/\sqrt{2}$. Assuming that $s$ subcarriers dedicated for user $U_k$, a logical channel $\mathcal{N}_{U_k}$ can be defined for the given AMQD block as

$$\mathcal{N}_{U_k}[j] = \left[\mathcal{N}_1, \ldots, \mathcal{N}_s\right]^T. \tag{30}$$

## 2.2 SVD on a Hilbert space

The SVD of an $M \times N$ real or complex matrix $\mathbf{M}$ is a factorization of $\mathbf{M}$, as follows:

$$\mathbf{M} = U \Sigma V^{-1}, \tag{31}$$

where $U$ is an $M \times M$ unitary matrix, $V^{-1}$ is an $N \times N$ unitary matrix, and $\Gamma$ is an $M \times N$ diagonal matrix [18-20]. The $\lambda_i$ nonnegative real diagonal entries of $\sum$ are the singular values of $\mathbf{M}$, and the columns of $U$ and $V^{-1}$ are referred as the left-singular vectors (eigenvectors of $\mathbf{MM}^\dagger$) and right-singular vectors of $\mathbf{M}$ (eigenvectors of $\mathbf{MM}^\dagger$). The $\lambda_i$ singular values are the square roots of the nonzero eigenvalues of $\mathbf{MM}^\dagger$ or $\mathbf{MM}^\dagger$.

In particular, the SVD of Equation (31) can be extended to a (separable) Hilbert space $\mathcal{H}$ via a bounded operator $\mathcal{B}(\cdot)$, using a partial isometry $U$, a unitary $V$, a measure space $Z$, and a nonnegative measureable $\mathcal{F}$, such that

$$\mathcal{B}(\mathbf{M}) = U\Gamma V^{-1}, \tag{32}$$

where $\Gamma$ is a multiplication by $\mathcal{F}$ on $Z^2$, which can be decomposed into

$$\mathcal{B}(\mathbf{M}) = UV^{-1} \cdot V\Gamma V^{-1}, \tag{33}$$

where $UV^{-1}$ identifies a partial isometry and $V\Gamma V^{-1}$ is positive.

## 3 Singular Layer for CVQKD

**Theorem 1** (Additional degree of freedom for AMQD via singular layer transmission). *The additional degree of freedom of the singular layer provides an improved $R$ transmission rate over the $l$ Gaussian sub-channels $\mathcal{N}_i$, upper bounded by $C_{AMQD}(\mathcal{N}) = \max_{\forall i} \sum_l \log_2 \left(1 + \frac{\sigma^2_{\omega''_i}|F(T_i(\mathcal{N}_i))|^2}{\sigma^2_\mathcal{N} + \sigma^2_\gamma}\right)$, where $\sigma^2_{\omega''} = \nu_{Eve} - \left(\sigma^2_\mathcal{N} \middle/ \max_{n_{\min}} \lambda^2_i\right)$, $\lambda_i$ is the $i$-th eigenchannel of $F(\mathbf{T})$, $\max_{n_{\min}} \lambda^2_i$ is the largest eigenvalue of $F(\mathbf{T})F(\mathbf{T})^\dagger$, and $\sigma^2_\gamma$ is the interference noise.*

*Proof.*
In terms of AMQD [4], the output $y_{k,m}$ refers to the $m$-th message of user $U_k$, and it is expressed as

$$y_{k,m} = F(\mathbf{T}(\mathcal{N}))z_{k,m} + F(\Delta). \tag{34}$$

The singular layer consists of a pre-unitary $F_1$ ($U_1$) (scaled FFT operation (scaled CVQFT), independent from the IFFT operation $F$ ($U_1$)) (see Figs. 1 and 2) and a post-unitary $U_2^{-1}$ (CVQFT operation, independent from the $U$ CVQFT$^\dagger$ operation) that perform the pre- and post-transform. The pre-unitary $F_1$ ($U_1$) transforms such that the input will be sent through the $\lambda_i$ eigenchannels of the Gaussian link, whereas $U_2^{-1}$ performs its inverse. Note that the pre-$F_1$ ($U_1$) and post-$U_2^{-1}$ unitaries are the not inverse of $F$ and $U$ but $F_1^{-1}$ ($U_1^{-1}$) and $U_2$, respectively.

In particular, these unitaries define the set $S_1$ of singular operators, as follows:

$$S_1 = \{F_1, U_2^{-1}\}. \tag{35}$$

Specifically, if each transmit user sends a single-carrier Gaussian CV signal to an encoder $\mathcal{E}$, then the pre-operator is the unitary $U_1$, the CVQFT operation, whereas the unitary post-operator is achieved by the inverse CVQFT operation $U_2^{-1}$, defining the set $S_2$ of singular operators as

$$S_2 = \{U_1, U_2^{-1}\}. \tag{36}$$

The subindices of the operators $\{F_1, U_2^{-1}\}$ and $\{U_1, U_2^{-1}\}$ are different in each $S_i, i = 1, 2$ because these operators are not the inverse of each other. These operators are determined by the SVD of $F(\mathbf{T})$, which is evaluated as

$$F(\mathbf{T}) = U_2 \Gamma F_1^{-1}, \tag{37}$$

where $F_1^{-1}, F_1 \in \mathbb{C}^{K_{in} \times K_{in}}$ and $U_2, U_2^{-1} \in \mathbb{C}^{K_{out} \times K_{out}}$, $K_{in}$, and $K_{out}$ refer to the number of sender and receiver users such that $K_{in} \leq K_{out}$, $F_1^{-1} F_1 = F_1 F_1^{-1} = I$, and $U_2 U_2^{-1} = U_2^{-1} U_2 = I$. The term $\Gamma \in \mathbb{R}$ is a diagonal matrix with nonnegative real diagonal elements

$$\lambda_1 \geq \lambda_2 \geq \ldots \lambda_{n_{\min}}, \tag{38}$$

which are called the *eigenchannels* of $F(\mathbf{T}) = U_2 \Gamma F_1^{-1}$, where

$$n_{\min} = \min(K_{in}, K_{out}), \tag{39}$$

which equals to the rank of $F(\mathbf{T})$, where

$$K_{in} \leq K_{out}, \tag{40}$$

by an initial assumption. (*Note*: the eigenchannels are also called the ordered singular values of $F(\mathbf{T})$.)

In terms of the $\lambda_i$ eigenchannels, $F(\mathbf{T})$ can be precisely rewritten as

$$F(\mathbf{T}) = \sum_{n_{\min}} \lambda_i U_{2,i} F_{1,i}^{-1}, \tag{41}$$

where $\lambda_i U_{2,i} F_{1,i}^{-1}$ are rank-one matrices.

In fact, the $n_{\min}$ squared eigenchannels $\lambda_i^2$ are the eigenvalues of the matrix

$$F(\mathbf{T}) F(\mathbf{T})^\dagger = U_2 \Gamma \Gamma^T U_2^{-1}, \tag{42}$$

where $\Gamma^T$ is the transpose of $\Gamma$.

The unitary $F_1(U_1)$ applied on the input data $\mathbf{z} \in \mathcal{CN}(0, \mathbf{K_z})$ defines an $\mathbf{s}$ stream matrix

$$\mathbf{s} = (s_1, \ldots, s_{n_{\min}})^T = F_1(\mathbf{z}) \in \mathcal{CN}(0, \mathbf{K_s}), \tag{43}$$

with $n_{\min}$ data streams $s_i$. A stream variable $s_i$ identifies the CV state $|s_i\rangle$ in the phase space $\mathcal{S}$ as

$$|s_i'\rangle = \lambda_i U_{2,i} F_{1,i}^{-1} |s_i\rangle. \tag{44}$$

The noisy $\mathbf{s}' = \mathcal{N}(\mathbf{s})$ stream vector is

$$\mathbf{s}' = (s_1', \ldots, s_{n_{\min}}')^T = F_1(\mathbf{z}') \in \mathcal{CN}(0, \mathbf{K_s} + \mathbf{K_\gamma} + \mathbf{K_\Delta}), \tag{45}$$

where $\mathbf{K}_\gamma$ is the covariance of the $\gamma \in \mathcal{CN}(0, \mathbf{K}_\gamma)$ eigenchannel interference and

$$|\mathbf{s}'\rangle = F(\mathbf{T})\mathbf{s} = U_2 \Gamma F_1^{-1} |\mathbf{s}\rangle \\ = \sum_{n_{\min}} \lambda_i U_{2,i} F_{1,i}^{-1} |s_i\rangle. \quad (46)$$

Without loss of generality, using the operations $F_1$, $F_1^{-1}$, and $U_2^{-1}$, Equation (34) can be rewritten as

$$U_2^{-1}(y_{k,m}) = F_1\left(\left(U_2 \Gamma F_1^{-1}\right) UF(z_{k,m})\right) U_2^{-1} + U_2^{-1}(U(\Delta) + \gamma_i) \\ = U_2 \Gamma F_1^{-1} z_{k,m} + U_2^{-1}(U(\Delta)) + U_2^{-1}(\gamma_i), \quad (47)$$

where $U_2$ and $U_2^{-1}$ unitaries stand for the CVQFT and inverse CVQFT operations, $U$ refers to the CVQFT operation, and $U_2^{-1}(U(\Delta))$ has the same distribution as $\Delta$,

$$U_2^{-1}(U(\Delta)) \in \mathcal{CN}(0, \mathbf{K}_\Delta), \quad (48)$$

whereas

$$U_2^{-1}(\gamma_i) \in \mathcal{CN}(0, \mathbf{K}_{\gamma_i}). \quad (49)$$

Thus, for simplicity, the term of $U_2^{-1}(U(\Delta))$ will be called $U(\Delta)$, and

$$\|F_1(\mathbf{z})\|^2 = \|\mathbf{z}\|^2, \quad (50)$$

by theory (see also the definitions of AMQD [4] and AMQD-MQA [3]). In case of AMQD, the subcarrier transmission is realized through the $l$ good $\mathcal{N}_i$ Gaussian sub-channels with a constant modulation variance $\sigma_\omega^2$. In the singular layer, the transmission is interpreted through the $n_{\min}$ eigenchannels $\lambda_i$, with a modulation variance $\sigma_{\omega'}^2 \neq \sigma_\omega^2$.

The unitary $F_1$ makes the transmission of data stream $s_i$ through the $\lambda_i$ eigenchannel possible. In particular, if the encoder $\mathcal{E}$ knows the SVD decomposition

$$F(\mathbf{T}) = U_2 \Gamma F_1^{-1}, \quad (51)$$

then it is possible for $\mathcal{E}$ to send the pre-transformed data streams $s_i$ through the $\lambda_i$ eigenchannels without interference $\gamma_i$. At a partial channel side information, the $\lambda_i$ eigenchannels of $\Gamma$ interfere with one another; thus, the streams will arrive non-orthogonally to the decoder $\mathcal{D}$, which requires an interference cancellation at $\mathcal{D}$.

The Fourier-transformed interference $|U_2^{-1}(\gamma_i)\rangle$, $U_2^{-1}(\gamma_i) \in \mathcal{CN}(0, \mathbf{K}_{\gamma_i})$, $\mathbf{K}_{\gamma_i} = \sigma_{\gamma_i}^2 = \left[\mathbb{E}|\gamma_i|^2\right]$, $\frac{1}{n_{\min}} \sum_{n_{\min}} \sigma_{\gamma_i}^2 = \sigma_\gamma^2$, is defined in the $\mathcal{S}$ phase space precisely as

$$|U_2^{-1}(\gamma_i)\rangle = U_2^{-1}\left(\sum_{j \neq i}^{n_{\min}} \lambda_j U_{2,j} F_{1,j}^{-1}\right) |s_j\rangle, \quad (52)$$

where $U_2^{-1}$ refers to the post-unitary inverse CVQFT operation. (*Note*: The $\gamma_i(\lambda_i)$ eigenchannel interference is not related to the crosstalk noise [4] $\gamma_i(\mathcal{N}_i)$, which could occur between the $\mathcal{N}_i$ Gaussian sub-channels during the transmission. In particular, the crosstalk noise is included in the noise variance $\sigma_\mathcal{N}^2$ of the Gaussian sub-channels; see [4].)

The unitary post-operation is performed by $U_2^{-1}$ at $\mathcal{D}$ on the noisy stream vector $\mathbf{s}'$ as
$$U_2^{-1}|\mathbf{s}'\rangle = |\mathbf{z}'\rangle = |\varphi\rangle. \tag{53}$$
In Fig. 1, the elements of the additional layer injected by the SVD are depicted above the functional components of AMQD-MQA.

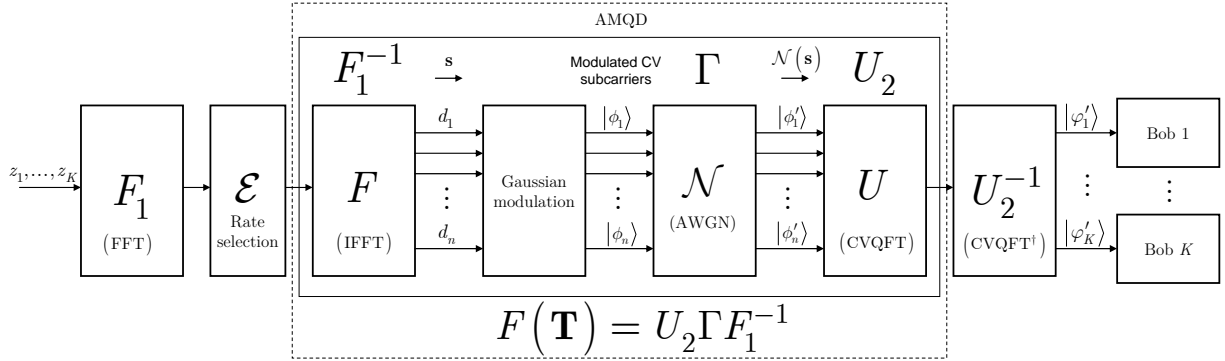

**Figure 1.** The singular layer transmission model of the $K \to K$ AMQD-MQA multiple-access scheme. The model consists of a single encoder and $K$-independent receivers. The SVD of AMQD is expressed by the matrix $F(\mathbf{T}) = U_2 \Gamma F_1^{-1}$. The pre- and post-unitaries of the singular layer are the $F_1$ scaled FFT operation and its unitary inverse CV operation $U_2^{-1}$.

In the second setting of AMQD-MQA, $F(\mathbf{T})$ is precisely evaluated as
$$F(\mathbf{T}) = U_2 \Gamma U_1^{-1}. \tag{54}$$
Thus, without loss of generality,
$$\begin{aligned} U_2^{-1}(y_{k,m}) &= U_1\left(\left(U_2 \Gamma F_1^{-1}\right) U F(z_{k,m})\right) U_2^{-1} + U_2^{-1}\left(U(\Delta) + \gamma_i\right) \\ &= U_2 \Gamma F_1^{-1} z_{k,m} + U_2^{-1}\left(U(\Delta)\right) + U_2^{-1}(\gamma_i). \end{aligned} \tag{55}$$
The SVD decomposition for the second setting of AMQD-MQA is depicted in Fig. 2.

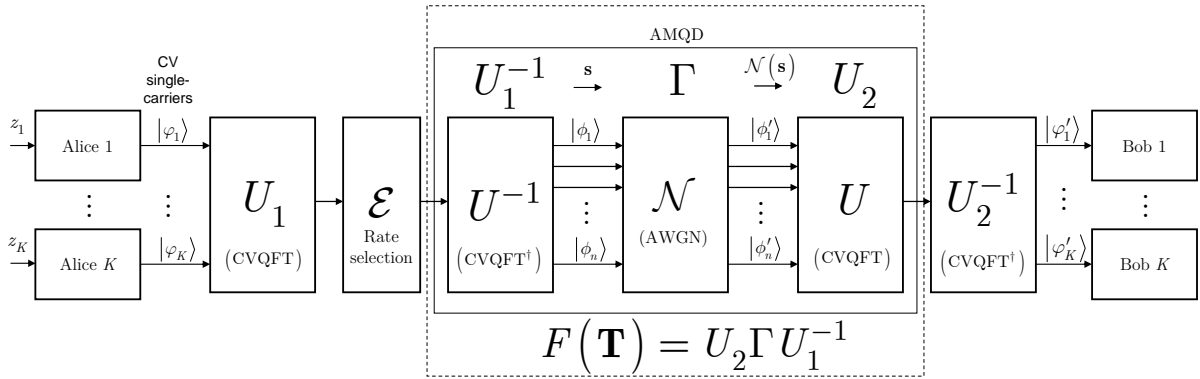

**Figure 2.** The singular layer transmission model of the $K \to K$ AMQD-MQA, with $K$-independent transmitters and $K$-independent receivers. The SVD of AMQD is expressed by the matrix $F(\mathbf{T}) = U_2 \Gamma U_1^{-1}$. The pre- and post-unitaries of the singular layer are the $U_1$ CVQFT operation and the inverse CVQFT operation $U_2^{-1}$.

In the AMQD framework, the optimal solution is to allocate the constant modulation variance $\sigma_{\omega'}^2$ for the $n_{\min}$ eigenchannel $\lambda_i$ because in the low-SNR regimes, it is provably the optimal solution [4]. Assuming the constant modulation variance $\sigma_\omega^2$ for the $l$ sub-channels

$$\sigma_{\omega'}^2 = \sum_{n_{\min}} \sigma_{\omega_i'}^2 = \frac{1}{l}\sum_l \sigma_{\omega_i}^2 = \sigma_\omega^2 \\ = \nu_{Eve} - \left(\sigma_\mathcal{N}^2 \bigg/ \max_{\forall i} |F(T_i(\mathcal{N}_i))|^2\right), \tag{56}$$

where $\max_{\forall i}|F(T_i(\mathcal{N}_i))|^2$ identifies the strongest Gaussian sub-channel $\mathcal{N}_i$ in AMQD, see [4], and

$$\sigma_\omega^2 = \tfrac{1}{1+c}\sigma_{\omega''}^2 = \tfrac{1}{1+c}\left[\nu_{Eve} - \left(\sigma_\mathcal{N}^2 \bigg/ \max_{n_{\min}} \lambda_i^2\right)\right], \tag{57}$$

where $c > 0$ is a constant, $\sigma_{\omega''}^2$ is the constant sub-channel modulation variance at a constant modulation variance $\sigma_{\omega'}^2$ for the $i$-th eigenchannel $\lambda_i$ in the singular layer, which is yielded as

$$\sigma_{\omega_i'}^2 = \mu - \left(\sigma_\mathcal{N}^2 \bigg/ \max_{n_{\min}} \lambda_i^2\right) \\ = \tfrac{1}{n_{\min}}\sigma_{\omega'}^2 \\ = \tfrac{1}{n_{\min}}\sigma_\omega^2 \\ = \tfrac{1}{n_{\min}}\tfrac{1}{1+c}\left[\nu_{Eve} - \left(\sigma_\mathcal{N}^2 \bigg/ \max_{n_{\min}} \lambda_i^2\right)\right], \tag{58}$$

where $\max_{n_{\min}} \lambda_i^2$ is the is the largest eigenvalue of $F(\mathbf{T})F(\mathbf{T})^\dagger$, thus $\mu$ is chosen as follows:

$$\mu = \tfrac{1}{n_{\min}(1+c)}\left[\nu_{Eve} + \left(n_{\min}(1+c)-1\right)\left(\sigma_\mathcal{N}^2 \bigg/ \max_{n_{\min}} \lambda_i^2\right)\right], \tag{59}$$

where $\nu_{Eve} = 1/\lambda$, and $\lambda$ is the Lagrange coefficient of Eve's optimal collective Gaussian attack, see [4]

From (58), the achievable $C(\mathcal{N})$ capacity of $\mathcal{N}$ over the $n_{\min}$ eigenchannels $\lambda_i$ is precisely:

$$C(\mathcal{N}) = \sum_{n_{\min}} \log_2\left(1 + \frac{\sigma_{\omega_i'}^2 \lambda_i^2}{\sigma_\mathcal{N}^2 + \sigma_\gamma^2}\right) \\ = \sum_{n_{\min}} \log_2\left(1 + \frac{\frac{1}{n_{\min}}\sigma_{\omega'}^2 \lambda_i^2}{\sigma_\mathcal{N}^2 + \sigma_\gamma^2}\right) = \sum_{n_{\min}} \log_2\left(1 + \frac{\sigma_{\omega'}^2 \lambda_i^2}{n_{\min}\left(\sigma_\mathcal{N}^2 + \sigma_\gamma^2\right)}\right), \tag{60}$$

where $\sigma_\gamma^2$ is the eigenchannel interference, which is vanishing $\sigma_\gamma^2 \to 0$, in an ideal scenario when the SVD of $F(\mathbf{T})$ is available for a precoding or postcoding at the transmitter or at the receiver, respectively. In particular, $\sigma_\gamma^2$ is also vanishing at the partial channel side information when only a statistical approximation $S(F(\mathbf{T}))$ of $F(\mathbf{T})$ is available for the parties (see Section 4), in which the phenomenon is much more significantly present (and also significant by practical purposes) in the low-SNR regimes, as it will be also proven in Section 4.

In particular, the singular layer transmission is allocated with the modulation variance $\sigma^2_{\omega'_i}$ of Equation (58) such that the constraint of Equation (56) is satisfied. At $n_{\min}$ eigenchannels, the covariance matrix $\mathbf{K}_{k,\mathbf{s}}$ of user $U_k$ is evaluated as follows:

$$\mathbf{K}_{k,\mathbf{s}} = F_1 diag\left\{\sigma^2_{\omega'_{k,1}},\ldots,\sigma^2_{\omega'_{k,n_{\min}}}\right\}F_1^{-1} \tag{61}$$

or

$$\mathbf{K}_{k,\mathbf{s}} = U_1 diag\left\{\sigma^2_{\omega'_{k,1}},\ldots,\sigma^2_{\omega'_{k,n_{\min}}}\right\}U_1^{-1}, \tag{62}$$

where the modulation variance $\sigma^2_{\omega'_{k,i}}$ is allocated to the $i$-th eigenchannel $\lambda_i$ of $U_k$, which leads to achievable capacity for $U_k$ as

$$\begin{aligned} C\left(\mathcal{N}^{U_k}\right) &= \log_2 \det\left(\mathbf{I}_{K_{out}} + \frac{1}{\sigma^2_{\mathcal{N}}} F(\mathbf{T}_k)\mathbf{K}_{k,\mathbf{s}} F(\mathbf{T}_k)^\dagger\right) \\ &= \sum_{n_{\min}} \log_2\left(1 + \mathrm{SNIR}_{k,i}\lambda^2_{k,i}\right), \end{aligned} \tag{63}$$

where SNIR is the signal-to-noise-plus-interference ratio.
The SNIR is precisely evaluated as

$$\mathrm{SNIR}_i = \frac{\sigma^2_{\omega'_i}}{\sigma^2_{\mathcal{N}}+\sigma^2_\gamma} = \frac{\frac{1}{n_{\min}}\sigma^2_\omega}{\sigma^2_{\mathcal{N}}+\sigma^2_\gamma}, \tag{64}$$

where $\sigma^2_\gamma$ is the variance of eigenchannel interference $U_2^{-1}(\gamma_i) \in \mathcal{CN}(0,\mathbf{K}_\gamma)$.

For $K_{in}$ transmit users, the resulting capacity is

$$\begin{aligned} C(\mathcal{N}) &= \log_2 \det\left(\mathbf{I}_{K_{out}} + \frac{1}{\sigma^2_{\mathcal{N}}} F(\mathbf{T})\mathbf{K}_{\mathbf{s}} F(\mathbf{T})^\dagger\right) \\ &= \log_2 \det\left(\mathbf{I}_{K_{out}} + \sum_{K_{in}}\frac{1}{\sigma^2_{\mathcal{N}}} F(\mathbf{T}_k)\mathbf{K}_{k,\mathbf{s}} F(\mathbf{T}_k)^\dagger\right) \\ &= \sum_{k=1}^{K_{in}}\sum_{n_{\min}} \log_2\left(1 + \mathrm{SNIR}_{k,i}\lambda^2_{k,i}\right), \end{aligned} \tag{65}$$

according to the capacity region of AMQD-MQA defined in [3].

In fact, assuming that the strongest eigenchannel is $\max_{n_{\min}} \lambda_i$ and all $\lambda_i$ eigenchannels pick up this maximum value in the low-SNR regimes ($\sigma^2_\gamma \to 0$), the achievable capacity is calculated as follows:

$$C(\mathcal{N}) \approx \sum_{n_{\min}} \max_{n_{\min}} \lambda^2_i \cdot \log_2 e \cdot \frac{\sigma^2_{\omega'_i}}{\sigma^2_{\mathcal{N}}}, \tag{66}$$

where

$$\max_{n_{\min}} \lambda^2_i \frac{\sum_{n_{\min}} \sigma^2_{\omega'_i}}{\sigma^2_{\mathcal{N}}} = \max_{n_{\min}} \lambda^2_i \frac{\sigma^2_\omega}{\sigma^2_{\mathcal{N}}}. \tag{67}$$

Thus,

$$\sum_{n_{\min}} \max_{n_{\min}} \lambda_i^2 > \max_{\forall i} \sum_l \left| F\left(T_i\left(\mathcal{N}_i\right)\right) \right|^2. \tag{68}$$

To see it, we exploit a property of AMQD. Let $0 \leq p \leq 1$ be the probability that the $\sum_l \left| F\left(T_i\left(\mathcal{N}_i\right)\right) \right|^2$ sum of the squared magnitude of the Fourier-transformed channel transmission coefficients of the $l$ Gaussian sub-channels pick up the maximum $\sum_l \left| F\left(T_i\left(\mathcal{N}_i\right)\right) \right|^2 = \max_{\forall i} \sum_l \left| F\left(T_i\left(\mathcal{N}_i\right)\right) \right|^2$:

$$p = \mathbb{E}\left[ \Pr\left( \sum_l \left| F\left(T_i\left(\mathcal{N}_i\right)\right) \right|^2 = \max_{\forall i} \sum_l \left| F\left(T_i\left(\mathcal{N}_i\right)\right) \right|^2 \right) \right]. \tag{69}$$

From this, the achievable rate over $\mathcal{N}$ is precisely

$$R_{AMQD}\left(\mathcal{N}\right) = p \log_2 \left( 1 + \mathrm{SNR} \, \frac{\max_i \sum_l \left| F\left(T_i\left(\mathcal{N}_i\right)\right) \right|^2}{p} \right), \tag{70}$$

where $\mathrm{SNR} = \frac{\sigma_\omega^2}{\sigma_\mathcal{N}^2}$, which can be further evaluated into

$$R_{AMQD}\left(\mathcal{N}\right) = \max_i \sum_l \left| F\left(T_i\left(\mathcal{N}_i\right)\right) \right|^2 \cdot \log_2 e \cdot \mathrm{SNR}. \tag{71}$$

Assuming that $\lambda_i$ eigenchannels are allocated with the modulation variance $\sigma_{\omega'}^2$, and assuming the probability that each squared eigenchannel $\lambda_i^2$ equals $\max_{n_{\min}} \lambda_i^2$,

$$p = \mathbb{E}\left[ \Pr\left( \sum_{n_{\min}} \lambda_i^2 = \sum_{n_{\min}} \max_{n_{\min}} \lambda_i^2 \right) \right], \tag{72}$$

one obtains rate, calculated as follows:

$$\begin{aligned} R_{\max_{n_{\min}} \lambda_i^2}\left(\mathcal{N}\right) &= p \log_2 \left( 1 + \frac{\sum_{n_{\min}} \max_{n_{\min}} \lambda_i^2}{p} \cdot \mathrm{SNR} \right) \\ &= \sum_{n_{\min}} \max_{n_{\min}} \lambda_i^2 \cdot \log_2 e \cdot \mathrm{SNR} \\ &= \sum_{n_{\min}} \max_{n_{\min}} \lambda_i^2 \cdot \log_2 e \cdot \frac{\sigma_{\omega_i'}^2}{\sigma_\mathcal{N}^2}, \end{aligned} \tag{73}$$

and from which Equation (66) is concluded.

As a conclusion, by performing the modulation variance allocation mechanism for the $l$ Gaussian sub-channels from [3-4] and by redefining $\nu_{\min} = \sigma_\mathcal{N}^2 \Big/ \max_{\forall i} \left| F\left(T_i\left(\mathcal{N}_i\right)\right) \right|^2$ from [4] as

$$\nu'_{\min} = \frac{\sigma_\mathcal{N}^2}{\max_{n_{\min}} \lambda_i^2}, \tag{74}$$

the constant modulation variance $\sigma_{\omega''}^2$ is derived as follows:

$$\begin{aligned} \sigma_{\omega''}^2 &= \nu_{Eve} - \nu'_{\min} \\ &= \nu_{Eve} - \frac{\sigma_\mathcal{N}^2}{\max_{n_{\min}} \lambda_i^2}, \end{aligned} \tag{75}$$

where $\nu_{Eve} = 1/\lambda$, and $\lambda$ is the Lagrange coefficient, with

$$\sigma^2_{\omega''} - \sigma^2_{\omega} = \nu_{\min} - \nu'_{\min} = \sigma^2_{\mathcal{N}} \left( \frac{1}{\max_{\forall i} |F(T_i(\mathcal{N}_i))|^2} - \frac{1}{\max_{n_{\min}} \lambda_i^2} \right). \tag{76}$$

The $C_{AMQD}(\mathcal{N})$ achievable capacity over the $l$ Gaussian sub-channels is increased in comparison with that of [4] to

$$C_{AMQD}(\mathcal{N}) = \max_{\forall i} \sum_{i=1}^{l} \log_2 \left( 1 + \frac{\sigma^2_{\omega''_i} |F(T_i(\mathcal{N}_i))|^2}{\sigma^2_{\mathcal{N}}} \right)$$

$$= \max_{\forall i} \sum_{i=1}^{l} \log_2 \left( 1 + \frac{\left[\nu_{Eve} - \frac{\sigma^2_{\mathcal{N}}}{\max_{n_{\min}} \lambda_i^2}\right] |F(T_i(\mathcal{N}_i))|^2}{\sigma^2_{\mathcal{N}}} \right), \tag{77}$$

with a total constraint

$$\tfrac{1}{l} \sum_l \sigma^2_{\omega''_i} = \sigma^2_{\omega''} = (1+c)\sigma^2_{\omega} > \sigma^2_{\omega}. \tag{78}$$

The higher-layer manipulations of AMQD can be exploited in a multicarrier scenario, using the $l$ Gaussian sub-channels for the transmission. In particular, the singular layer leads to improved capacity and improved user rates in the multiuser scheme of AMQD-MQA. However, the modulation variance allocation mechanism requires some refinements in comparison with that of [4].

The algorithm for the enhanced modulation variance allocation is precisely as follows.

**Algorithm**
1. *Perform the singular layer transmission via the unitary pre- and post operators $F_1$, $U_2^{-1}$, or $U_1, U_2^{-1}$, where $F_1$ is the scaled FFT operation, $U_1$ is the CVQFT operation, and $U_2^{-1}$ is the inverse CVQFT, respectively. Compute $F(\mathbf{T}) = U_2 \Gamma F_1^{-1}$ ($F(\mathbf{T}) = U_2 \Gamma U_1^{-1}$), where $\Gamma \in \mathbb{R}$ is a diagonal matrix with non-negative real diagonal elements $\lambda_1 \geq \lambda_2 \geq \ldots \lambda_{n_{\min}}$, where $n_{\min} = \min(K_{in}, K_{out})$, at $K_{in}$ transmit and $K_{out}$ receiver users and $l$ good Gaussian sub-channels.*
2. *Determine the $n_{\min}$ squared eigenchannels $\lambda_i^2$ from $\Gamma$ or from the eigenvalues of the matrix $F(\mathbf{T}) F(\mathbf{T})^\dagger = U_2 \Gamma \Gamma^T U_2^{-1}$, where $\Gamma^T$ is the transpose of $\Gamma$.*
3. *Determine $\max_{n_{\min}} \lambda_i^2$, where $\sum_{n_{\min}} \max_{n_{\min}} \lambda_i^2 > \max_{\forall i} \sum_l |F(T_i(\mathcal{N}_i))|^2$, and compute the constant modulation variance $\sigma^2_{\omega'}$ for the $\lambda_i$ ei-*

*genchannels as* $\sigma_{\omega'_i}^2 = \mu - \left(\sigma_{\mathcal{N}}^2 \big/ \max\limits_{n_{\min}} \lambda_i^2\right)$, *where* $\mu$ *is chosen as*

$$\mu = \tfrac{1}{n_{\min}(1+c)} \left(\nu_{Eve} + \left(n_{\min}(1+c) - 1\right)\left(\sigma_{\mathcal{N}}^2 \big/ \max\limits_{n_{\min}} \lambda_i^2\right)\right), \text{ with a total}$$

*constraint* $\sigma_{\omega'}^2 = \sum_{n_{\min}} \sigma_{\omega'_i}^2 = \tfrac{1}{l}\sum_l \sigma_{\omega_i}^2 = \sigma_\omega^2$.

4. *Determine SNIR as* $\frac{\sum_{n_{\min}} \sigma_{\omega'_i}^2}{\sigma_{\mathcal{N}}^2 + \sigma_\gamma^2} = \frac{\sigma_\omega^2}{\sigma_{\mathcal{N}}^2 + \sigma_\gamma^2}$. *The achievable rate in the low-SNR regimes is* $R(\mathcal{N}) \leq \sum_{n_{\min}} \max\limits_{n_{\min}} \lambda_i^2 \cdot \log_2 e \cdot \frac{\sigma_{\omega'_i}^2}{\sigma_{\mathcal{N}}^2}$, $\sigma_\gamma^2 \to 0$, *if all* $\lambda_i$ *eigenchannels pick up the theoretical maximum* $\max\limits_{n_{\min}} \lambda_i^2$.

5. *Perform the allocation of the l Gaussian sub-channels with the modulation variance* $\sigma_{\omega''_i}^2 = \nu_{Eve} - \nu'_{\min} = \nu_{Eve} - \left(\sigma_{\mathcal{N}}^2 \big/ \max\limits_{n_{\min}} \lambda_i^2\right)$, *where* $\nu'_{\min} = \sigma_{\mathcal{N}}^2 \big/ \max\limits_{n_{\min}} \lambda_i^2$, $\nu_{Eve} = 1/\lambda$, *and* $\lambda$ *is the Lagrange coefficient. Then* $R_{AMQD}(\mathcal{N}) = \max\limits_{\forall i} \sum_{i=1}^{l} \log_2 \left(1 + \frac{\sigma_{\omega''_i}^2 |F(T_i(\mathcal{N}_i))|^2}{\sigma_{\mathcal{N}}^2 + \sigma_\gamma^2}\right)$ *over the Gaussian sub-channels, where* $\sigma_\gamma^2 \to 0$ *in the low-SNR regimes.*

6. *If* $\Pi < \nu_\kappa$, *perform the compensation of a nonideal Gaussian modulation by* $\nu'_\kappa = \nu_\kappa - \Pi$ *for the l Gaussian sub-channels, where* $\Pi = \nu_{\min} - \nu'_{\min} = \sigma_{\mathcal{N}}^2 \big/ \max\limits_{\forall i} |F(T_i(\mathcal{N}_i))|^2 - \sigma_{\mathcal{N}}^2 \big/ \max\limits_{n_{\min}} \lambda_i^2$ *and* $\nu_\kappa$ *is the constant correction term of the noise level of the sub-channels.*

Note that in step 5, the allocation mechanism of the Gaussian sub-channels can be found in [4], and in step 6, the method for the compensation of the imperfections of a nonideal Gaussian input modulation is handled by the proposed algorithm of [3].

The results, in fact, conclude that from the SVD of $F(\mathbf{T}) = U_2 \Gamma F_1^{-1}$ and $F(\mathbf{T}) = U_2 \Gamma U_1^{-1}$, the $n_{\min}$ additional degree of freedom can be exploited to reach an improved performance in the multicarrier transmission of AMQD.

∎

The effect of the singular layer on the modulation variance allocation mechanism of the Gaussian sub-channels is illustrated in Fig. 3.

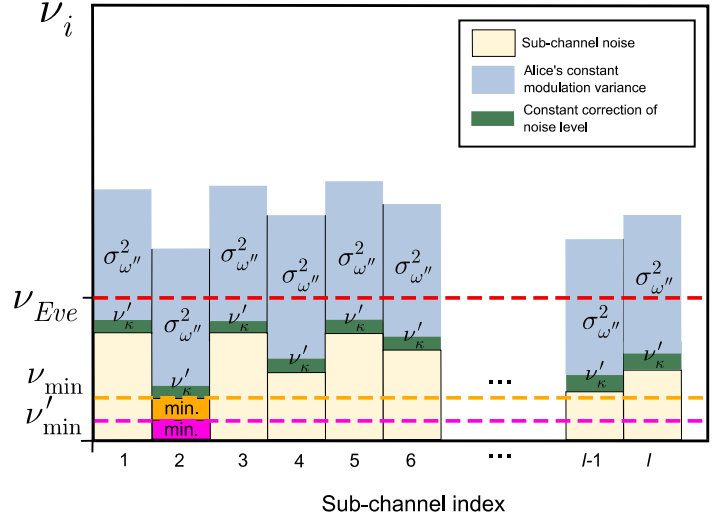

**Figure 3.** The effect of singular layer on the sub-channel modulation variance. In an AMQD modulation, the Gaussian subcarrier CVs are transmitted over $l$ good (i.e., $\nu_i < \nu_{Eve}$) Gaussian sub-channels $\mathcal{N}_i$. The sub-channel noise $\nu_{\min}$ is decreased from $\nu_{\min} = \sigma_{\mathcal{N}}^2 \Big/ \max_{\forall i} \left| F\left(T_i\left(\mathcal{N}_i\right)\right) \right|^2$ to $\nu'_{\min} = \sigma_{\mathcal{N}}^2 \Big/ \max_{n_{\min}} \lambda_i^2$, which results in an increased modulation variance $\sigma_{\omega''}^2 = \nu_{Eve} - \nu'_{\min}$ and a decreased constant correction $\nu'_\kappa = \nu_\kappa - \Pi$ in the noise level. The total constraint is increased to $\frac{1}{l}\sum_l \sigma_{\omega_i''}^2 = \sigma_{\omega''}^2 > \sigma_\omega^2$.

## 4 SIA at a Partial Channel Side Information

### 4.1 Transmission at a partial channel side information

**Lemma 1**. *At a partial channel side information, there exists an $F(\mathbf{T})$-independent pre-unitary operation $Q$ such that $F(\mathbf{T}) = \xi_{K_{out}} S(F(\mathbf{T})) \xi_{K_{in}}^{-1}$, where $S(F(\mathbf{T})) = \xi_{K_{out}}^{-1} \Gamma \xi_{K_{in}}$ is the statistical model of $F(\mathbf{T})$, $\xi_{K_{out}}^{-1}$ and $\xi_{K_{in}}$ are unitary operators that formulate the input covariance matrix $\mathbf{K_s} = \xi_{K_{in}} \wp \xi_{K_{in}}^{-1}$, and $\wp$ is a diagonal matrix. At $S(F(\mathbf{T}))$, $\sigma_\gamma^2 \to 0$ in the low-SNR regimes.*

*Proof.*
At a partial channel side information, the exact SVD of $F(\mathbf{T})$ is not available for the encoder; thus, the eigenvectors of $F(\mathbf{T})F(\mathbf{T})^\dagger$ are not available for the pre-unitary operation $F_1$ and $U_1$. In this case, the encoder $\mathcal{E}$ can use an $F(\mathbf{T})$-*independent* pre-unitary $Q$. In particular, the transmission is allocated with the modulation variance of Equation (58) such that the constraint of Equation (56) is satisfied, which at $K_{in}$ transmit users and input vector $\mathbf{s} = \left(s_1, \ldots, s_{K_{in}}\right)^T$, leading to the covariance matrix

$$\mathbf{K}_\mathbf{s} = Q diag\left\{\sigma^2_{\omega'_1},\ldots,\sigma^2_{\omega'_{K_{in}}}\right\}Q^\dagger, \tag{79}$$

and the $R$ reliable transmission rate is

$$R(\mathcal{N}) \leq \mathbb{E}\left[\log_2 \det\left(\mathbf{I}_{K_{out}} + \frac{1}{\sigma^2_\mathcal{N}} F(\mathbf{T})\mathbf{K}_\mathbf{s} F(\mathbf{T})^\dagger\right)\right], \tag{80}$$

which can be simply verified by a sphere packing argument [18-20].

Using Equation (79) with the assumption $K_{out} = 1$ and $Q = \mathbf{I}_{K_{in}}$, the maximal rate leads to Equation (60). Thus, the capacity of the Gaussian quantum channel can be achieved at an $F(\mathbf{T})$-independent unitary $Q$ because the covariance matrix $\mathbf{K}_\mathbf{s}$ in Equation (79) can be chosen as a function $f(\cdot)$ of the distribution of $F(\mathbf{T})$.

At a partial channel side information, $\mathbf{K}_\mathbf{s}$ requires the transmitter to build a statistical model $S(F(\mathbf{T})) = f(F(\mathbf{T}))$ about the distribution of $F(\mathbf{T})$, which allows to reach $\sigma^2_\gamma \to 0$ in the low-SNR regimes. It then leads to a reliable transmission rate of

$$\begin{aligned}\max R(\mathcal{N}) &= \max_{\mathbf{K}_\mathbf{s}: Tr(\mathbf{K}_\mathbf{s}) \leq \sigma^2_\omega} \mathbb{E}\left[\log_2 \det\left(\mathbf{I}_{K_{out}} + \frac{1}{\sigma^2_\mathcal{N}} F(\mathbf{T})\mathbf{K}_\mathbf{s} F(\mathbf{T})^\dagger\right)\right] \\ &= \sum_{n_{\min}} \mathbb{E}\left[\log_2\left(1 + \mathrm{SNR}_i \lambda^2_i\right)\right].\end{aligned} \tag{81}$$

The $\mathbf{K}_\mathbf{s}$ of the input has to be chosen such that it matches $S(F(\mathbf{T}))$ rather than the exact realization of the distribution of $F(\mathbf{T})$ [18] because at a partial channel side information, only $S(F(\mathbf{T}))$ is available for Alice. In particular, by exploiting the fact that Eve attacks all of the $\mathcal{N}_i$ Gaussian sub-channels, and the distribution of $F(\mathbf{T})$ formulates a time-independent deterministic process, $\mathcal{E}$ can choose an $F(\mathbf{T})$-independent $Q$ such that

$$Q = S(F_1) \text{ or } Q = S(U_1). \tag{82}$$

Thus, $S(F(\mathbf{T}))$ allows for Alice to use

$$S(F(\mathbf{T})) = \mathbb{E}\left[F_2 \Gamma F_1^{-1}\right] \tag{83}$$

or

$$S(F(\mathbf{T})) = \mathbb{E}\left[U_2 \Gamma U_1^{-1}\right]. \tag{84}$$

The results in Equations (83) and (84) allow for the transmitter to determine the eigenvalues of

$$S(F(\mathbf{T}))S(F(\mathbf{T}))^\dagger = \mathbb{E}\left[F_2 \Gamma \Gamma^T F_2^{-1}\right] \tag{85}$$

or

$$S(F(\mathbf{T}))S(F(\mathbf{T}))^\dagger = \mathbb{E}\left[U_2 \Gamma \Gamma^T U_2^{-1}\right], \tag{86}$$

In fact, operation $Q$ transforms the input to transmit them over the eigendirections of $F(\mathbf{T})^\dagger F(\mathbf{T})$; thus, the allocation algorithm of Theorem 1 can be directly applied in the input, which precisely leads to the maximal rate of Equation (77).

In particular, if the distribution of $F(\mathbf{T})$ cannot be modeled as a deterministic process, the Alice still can handle the situation via $S(F(\mathbf{T}))$ as follows. In this case, the entries of matrix $S(F(\mathbf{T}))$ are statistically independent with a zero mean.

Thus, it formulates another (i.e., different from the SVD of $F(\mathbf{T})$) matrix as

$$S(F(\mathbf{T})) = \xi_{K_{out}}^{-1} \Gamma \xi_{K_{in}}, \tag{87}$$

where $\xi_{K_{out}}^{-1}$ and $\xi_{K_{in}}$ *are* unitary operators. Thus,

$$F(\mathbf{T}) = \xi_{K_{out}} S(F(\mathbf{T})) \xi_{K_{in}}^{-1}, \tag{88}$$

and the pre-unitary $Q$ is

$$Q = \xi_{K_{in}}, \tag{89}$$

which leads to the $\mathbf{K_s}$ covariance matrix

$$\mathbf{K_s} = \xi_{K_{in}} \wp \xi_{K_{in}}^{-1}, \tag{90}$$

where $\wp$ is a diagonal matrix that identifies the $\sigma_{\omega'}^2$ modulation variance components.

Without loss of generality, the $S(F(\mathbf{T})) = \xi_{K_{out}}^{-1} \Gamma \xi_{K_{in}}$, and $\mathbf{K_s}$ leads to achievable capacity

$$\begin{aligned}
C(\mathcal{N}) &= \max_{\mathbf{K_s}:Tr(\mathbf{K_s})\leq \sigma_\omega^2} \mathbb{E}\left[\log_2 \det\left(\mathbf{I}_{K_{out}} + \tfrac{1}{\sigma_\mathcal{N}^2} S(F(\mathbf{T})) \xi_{K_{in}}^{-1} \mathbf{K_s} \xi_{K_{in}} S(F(\mathbf{T}))^\dagger\right)\right] \\
&= \max_{\mathbf{K_s}:Tr(\mathbf{K_s})\leq \sigma_\omega^2} \mathbb{E}\left[\log_2 \det\left(\mathbf{I}_{K_{out}} + \tfrac{1}{\sigma_\mathcal{N}^2} S(F(\mathbf{T})) \xi_{K_{in}}^{-1} \xi_{K_{in}} \wp \xi_{K_{in}}^{-1} \xi_{K_{in}} S(F(\mathbf{T}))^\dagger\right)\right],
\end{aligned} \tag{91}$$

where the diagonal entries of $\wp$ are chosen such that the optimality condition

$$\wp = \sigma_{\omega'}^2 \tfrac{1}{K_{in}} \mathbf{I}_{K_{in}} \tag{92}$$

is met, where $\sigma_{\omega'}^2 = \sum_{n_{\min}} \sigma_{\omega'_i}^2$ with the constraint of Equation (56).

Putting the pieces together, (91) leads to

$$\begin{aligned}
C(\mathcal{N}) &= \max_{\wp:Tr(\wp)\leq \sigma_\omega^2} \mathbb{E}\left[\log_2 \det\left(\mathbf{I}_{K_{out}} + \tfrac{1}{\sigma_\mathcal{N}^2} \xi_{K_{out}} S(F(\mathbf{T})) \wp S(F(\mathbf{T}))^\dagger \xi_{K_{out}}^{-1}\right)\right] \\
&= \max_{\wp:Tr(\wp)\leq \sigma_\omega^2} \mathbb{E}\left[\log_2 \det\left(\mathbf{I}_{K_{out}} + \tfrac{1}{\sigma_\mathcal{N}^2} S(F(\mathbf{T})) \wp S(F(\mathbf{T}))^\dagger\right)\right] \\
&= \mathbb{E}\left[\log_2 \det\left(\mathbf{I}_{K_{out}} + \tfrac{\frac{1}{K_{in}n_{\min}}\sigma_{\omega'}^2}{\sigma_\mathcal{N}^2 K_{in}} F(\mathbf{T}) F(\mathbf{T})^\dagger\right)\right] \\
&= \mathbb{E}\left[\sum_{n_{\min}} \log_2\left(1 + \tfrac{\frac{1}{K_{in}n_{\min}}\sigma_{\omega'}^2}{\sigma_\mathcal{N}^2} \lambda_i^2\right)\right] \\
&= \sum_{n_{\min}} \mathbb{E}\left[\log_2\left(1 + \tfrac{\frac{1}{K_{in}n_{\min}}\sigma_{\omega'}^2}{\sigma_\mathcal{N}^2} \lambda_i^2\right)\right],
\end{aligned} \tag{93}$$

where $\lambda_i^2$ are the ordered squared, random singular values of $F(\mathbf{T})$.

For the random distribution of $\lambda_i^2$, from the Jensen inequality [18], it precisely follows that

$$\sum_{n_{\min}} \mathbb{E}\left[\log_2\left(1 + \frac{\frac{1}{K_{in}n_{\min}}\sigma_{\omega'}^2}{\sigma_{\mathcal{N}}^2}\lambda_i^2\right)\right] \leq n_{\min}\mathbb{E}\left[\log_2\left(1 + \frac{\frac{1}{K_{in}n_{\min}}\sigma_{\omega'}^2}{\sigma_{\mathcal{N}}^2}\left(\frac{1}{n_{\min}}\sum_{n_{\min}}\lambda_i^2\right)\right)\right], \tag{94}$$

which demonstrates that Equation (93) is maximized if $S(F(\mathbf{T}))$ identifies a random process, in fact, if the distribution of $\lambda_i^2$ converges into a $\mathcal{U}$ uniform distribution.

In particular, in long-distance CVQKD scenarios, the transmission is realized in the low-SNR regimes, and we focus specifically on this segment. In the low-SNR regime, the term of $\frac{\frac{1}{K_{in}n_{\min}}\sigma_{\omega'}^2}{\sigma_{\mathcal{N}}^2}\lambda_i^2$ allows us to rewrite Equation (93) as

$$\begin{aligned}C(\mathcal{N}) &= \sum_{n_{\min}} \mathbb{E}\left[\log_2\left(1 + \frac{\frac{1}{K_{in}n_{\min}}\sigma_{\omega'}^2}{\sigma_{\mathcal{N}}^2}\lambda_i^2\right)\right]\\ &= \sum_{n_{\min}} \frac{\frac{1}{K_{in}n_{\min}}\sigma_{\omega'}^2}{\sigma_{\mathcal{N}}^2}\mathbb{E}\left[\lambda_i^2\right]\log_2 e\\ &= \frac{\frac{1}{K_{in}n_{\min}}\sigma_{\omega'}^2}{\sigma_{\mathcal{N}}^2}\mathbb{E}\left[Tr\left(F(\mathbf{T})F(\mathbf{T})^{\dagger}\right)\right]\log_2 e\\ &\quad \frac{\frac{1}{n_{\min}}\sigma_{\omega'}^2}{\sigma_{\mathcal{N}}^2}\frac{(K_{in}\cdot K_{out})}{K_{in}}\log_2 e\\ &= K_{out}\frac{\frac{1}{n_{\min}}\sigma_{\omega'}^2}{\sigma_{\mathcal{N}}^2}\log_2 e,\end{aligned} \tag{95}$$

where $\log_2\left(1 + \frac{\frac{1}{K_{in}n_{\min}}\sigma_{\omega'}^2}{\sigma_{\mathcal{N}}^2}\lambda_i^2\right) \approx \frac{\frac{1}{K_{in}n_{\min}}\sigma_{\omega'}^2}{\sigma_{\mathcal{N}}^2}\lambda_i^2 \log_2 e$, by theory. The result in Equation (95) reveals that in the low-SNR regimes, the distribution of $\lambda_i^2$ has no relevance [18] because $\mathbb{E}\left[\lambda_i^2\right]$ does not deal with the correlation among $\lambda_i^2$; and, the inequality of Equation (94) vanishes in the low-SNR CVQKD scenarios.

Note, that assuming that $K_{in} = K_{out} = K$, from the law of large numbers the capacity at $S(F(\mathbf{T}))$ can be approximated as

$$C(\mathcal{N}) = Kc \cdot \mathbb{E}[\text{SNR}], \tag{96}$$

where $c$ is a constant.

∎

## 4.2 Postcoding interference cancellation

**Lemma 2** (Interference cancellation at the decoder). *If $S(F(\mathbf{T})) = \mathbb{E}\left[U_2\Gamma F_1^{-1}\right]$ is available at the decoder, then the interference $\left|U_2^{-1}(\gamma_i)\right\rangle = U_2^{-1}\left|\gamma_i\right\rangle$ can be cancelled by operator $\mathcal{P}_i$ as $\mathcal{P}_i(\left|s_i'\right\rangle) = \amalg_i^{\perp}$, where $\amalg_i^{\perp} = \left\{\left|s_j'\right\rangle\right\}_{j=1}^{n_{\min}}$, $j \neq i$ consists of $n_{\min}$ CV states $\left|s_j'\right\rangle$, such that each $\left|s_j'\right\rangle$ is orthogonal to the subspace $\amalg_i$ spanned by the $n_{\min} - 1$ eigenchannel vectors $\left\{\lambda_j U_{2,j} F_{1,j}^{-1}\right\}$, $j = 1, \ldots, n_{\min}, j \neq i$.*

*Proof.*

Let the noisy output CV stream $|s_i'\rangle \in \mathcal{S}$ to be given as

$$|s_i'\rangle = |s_i\rangle + U_2^{-1}|\gamma_i\rangle + U|\Delta_i\rangle, \qquad (97)$$

where $U|\Delta_i\rangle = |U(\Delta_i)\rangle$ is the CVQFT-transformed noise of the Gaussian channel $\mathcal{N}$ (for further information see [4]), and $U_2^{-1}|\gamma_i\rangle$ is the inverse CVQFT-transformed interference of the $\lambda_i, i = 1,\ldots,n_{\min}$ eigenchannels, evaluated as

$$U_2^{-1}|\gamma_i\rangle = \mathbb{E}\left[U_2^{-1}\left(\sum_{j\neq i}^{n_{\min}} \lambda_j U_{2,j} F_{1,j}^{-1}\right)\right]|s_j\rangle. \qquad (98)$$

Assuming that the decoder knows $S(F(\mathbf{T})) = \mathbb{E}[U_2 \Gamma F_1^{-1}]$, it is possible to construct a $\mathcal{P}_i$ operator such that

$$\mathcal{P}_i(|s_i'\rangle) = \mathrm{II}_i^{\perp}, \qquad (99)$$

where the set

$$\mathrm{II}_i^{\perp} = \{|s_j'\rangle\}_{j=1}^{n_{\min}}, \; j \neq i \qquad (100)$$

consists of $n_{\min}$ CV states $|s_j'\rangle$, such that each $|s_j'\rangle$ is orthogonal to the subspace $\mathrm{II}_i$ spanned by the $n_{\min} - 1$ eigenchannel vectors, calculated as follows:

$$\{\lambda_j U_{2,j} F_{1,j}^{-1}\}, \; j = 1,\ldots,n_{\min}, j \neq i. \qquad (101)$$

In particular, operator $\mathcal{P}_i$ defines a $d_{S_i^{\perp}} \times K_{out}$ matrix, and the $i$-th row of $\mathcal{P}_i$ form an orthonormal basis $b_i$ of $\mathrm{II}_i^{\perp}$. Without loss of generality, the $d_{S_i^{\perp}}$ rows of $\mathrm{II}_i^{\perp}$,

$$b = \left\{b_1,\ldots,b_{d_{S_i^{\perp}}}\right\}, \qquad (102)$$

are all orthogonal to the subspace $\mathrm{II}_i$ spanned by the $n_{\min} - 1$ eigenchannel vectors $\{\lambda_j U_{2,j} F_{1,j}^{-1}\}, \; j \neq i$.

Specifically, if the SVD of $S(F(\mathbf{T})) = \mathbb{E}[U_2 \Gamma F_1^{-1}]$ is not the linear combination of any other SVDs of $F(T_j)$, $j \neq i$, then $U_2^{-1}|\gamma_i\rangle$ can be cancelled by applying $\mathcal{P}_i$ on $|s_j'\rangle$, resulting in

$$\mathcal{P}_i|s_i'\rangle = \mathcal{P}_i \lambda_i U_{2,i} F_{1,i}^{-1}|s_i\rangle + \mathcal{P}_i U_2^{-1}|\gamma_i\rangle + \mathcal{P}_i U|\Delta_i\rangle, \qquad (103)$$

where $\mathcal{P}_i U_2^{-1}|\gamma_i\rangle = \varnothing$, and $\mathcal{P}_i U|\Delta_i\rangle$ identifies a Gaussian noise vector $\mathcal{P}_i U|\Delta_i\rangle$ in the phase space with a distribution

$$\mathcal{P}_i U|\Delta_i\rangle \in \mathcal{CN}(0, \mathbf{K}_{\Delta_i}). \qquad (104)$$

In particular, the noise given in Equation (104) then leads to SNR, calculated as follows:

$$\mathrm{SNR} = \mathbb{E}\left[\frac{\sigma_{\omega''}^2 |\mathcal{P}_i(F(T_i(\mathcal{N}_i)))|^2}{\sigma_{\mathcal{N}}^2}\right]. \qquad (105)$$

The decoder-side interference cancellation can be rewritten by an operator $W_i$, for all $i$, as

$$W_i = \mathcal{P}_i^\dagger \mathcal{P}_i \left(\lambda_i U_{2,i} F_{1,i}^{-1}\right). \tag{106}$$

Let $-I$ be the pseudoinverse operation [18-20]. Because for the SVD of $F(T_i)$, the relation

$$\begin{aligned}\left(\lambda_i U_{2,i} F_{1,i}^{-1}\right)^{-I} &= \left[\left(\lambda_i U_{2,i} F_{1,i}^{-1}\right)^\dagger \left(\lambda_i U_{2,i} F_{1,i}^{-1}\right)\right]^{-1} \left(\lambda_i U_{2,i} F_{1,i}^{-1}\right)^\dagger \\ &= \left(\lambda_i U_{2,i} F_{1,i}^{-1}\right)^{-1}\end{aligned} \tag{107}$$

holds because $F(\mathbf{T})$ is a square matrix and invertible.

It leads to the conclusion that $\left(\lambda_i U_{2,i} F_{1,i}^{-1}\right)^\dagger \lambda_i U_{2,i} F_{1,i}^{-1} = I$ is invertible; thus, $\lambda_i U_{2,i} F_{1,i}^{-1}$ is linearly independent (i.e., it cannot be expressed as the linear combination of any other SVDs $\left(\lambda_j U_{2,j} F_{1,j}^{-1}\right)$, $j \neq i$), for $\forall i$.

Thus, the overall stream-level rate at the utilization of projector $\mathcal{P}_i$ is

$$R_{stream}(\lambda) \leq \mathbb{E}\left(\sum_{n_{\min}} \log_2\left(1 + \frac{\sigma_{\omega'}^2 \lambda_i^2}{\sigma_\mathcal{N}^2}\right)\right). \tag{108}$$

These results conclude the proof on the decoder-side eigenchannel interference cancellation.

∎

Note that the application of operator $W_i$ requires the knowledge of $S(F(\mathbf{T})) = \mathbb{E}\left[U_2 \Gamma F_1^{-1}\right]$ at the decoder. Thus, this type of interference cancellation is suboptimal. This problem will be resolved in Theorem 2.

## 4.3 SIA precoding

**Theorem 2** (SIA precoding of Gaussian random quadratures). *The $U_2^{-1}|\gamma_i\rangle$ eigenchannel interference can be cancelled by a precoding operator $V$ at encoder $\mathcal{E}$, which defines a phase space constellation $\mathcal{C}_\mathcal{S}$ such that the transmitted $|\tilde{\kappa}_i\rangle = \min\left\{EC_k|\varphi_i\rangle - \alpha U_2^{-1}|\gamma_i\rangle\right\}$ is the minimal difference of the equivalence class $EC_k|\varphi_i\rangle$ of the Gaussian CV $|\varphi_i\rangle$ and the interference term $\alpha U_2^{-1}|\gamma_i\rangle$, where $\alpha = \sigma_\omega^2 / \sigma_\omega^2 + \sigma_\mathcal{N}^2$. Operator $V$ is optimal at $S(F(\mathbf{T}))$ for all settings of AMQD-MQA.*

*Proof.*
At a $S(F(\mathbf{T}))$ partial channel side information, the streams that sent over the eigenchannels cannot be perfectly separated orthogonally by $F_1$ (scaled FFT operation) and $U_1$. Assuming an interference-free transmission, the decoder receives the noisy CV state $|\varphi_i'\rangle = |\varphi_i\rangle + U|\Delta_i\rangle$. In particular, in the AMQD setting, the inference is Fourier-transformed by the receiver, denoted by $U|\gamma_i\rangle$. Thus, the noisy Gaussian CV $|\varphi_i'\rangle$ is $|\varphi_i'\rangle = |\varphi_i\rangle + U_2^{-1}|\gamma_i\rangle + U|\Delta_i\rangle$.

As illustrated in Fig. 4, the eigenchannel interference $U_2^{-1}(\gamma_i)$ is an additional noise in the phase space $\mathcal{S}$, added to the transmit user Gaussian CV state $|\varphi_i\rangle$.

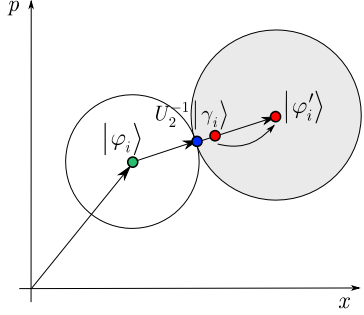

**Figure 4.** At eigenchannel interference, the user signal $|\varphi_i\rangle$ is transformed into $|\varphi_i\rangle + U_2^{-1}|\gamma_i\rangle$, which is then distorted into $|\varphi_i'\rangle = |\varphi_i\rangle + U_2^{-1}|\gamma_i\rangle + U|\Delta_i\rangle$. With no interference, the decoder receives $|\varphi_i'\rangle = |\varphi_i\rangle + U|\Delta_i\rangle$ ($x$, position quadrature; $p$, momentum quadrature).

Assume that a constellation $\mathcal{C}_\mathcal{S}$ in $\mathcal{S}$ exists and user $U_i$ transmits a Gaussian CV $|\varphi_i\rangle$ of $\mathcal{C}_\mathcal{S}$. In the naive approach, the $\mathcal{E}$ encoder compensates $U_2^{-1}|\gamma_i\rangle$ as follows. It encodes $|\varphi_i\rangle$ as

$$|\kappa_i\rangle = |\varphi_i\rangle + \left(-U_2^{-1}|\gamma_i\rangle\right) = |\varphi_i\rangle - U_2^{-1}|\gamma_i\rangle. \tag{109}$$

The CV state $|\kappa_i\rangle$ is then sent to the receiver, which results in the noisy $|\kappa_i'\rangle$ as follows:

$$|\kappa_i'\rangle = |\kappa_i\rangle + U_2^{-1}|\gamma_i\rangle + U|\Delta_i\rangle = |\varphi_i\rangle + U|\Delta_i\rangle. \tag{110}$$

In particular, the naive precoding approach is strictly suboptimal because the energy increase $\sigma_{\gamma_i}^2$ required in Equation (109) is unbounded by theory [18-20], because the transmit user Gaussian CV $|\varphi_i\rangle$ and the inference $U_2^{-1}|\gamma_i\rangle$ can be arbitrary distant from each other in $\mathcal{S}$. Thus, the compensation of $U_2^{-1}|\gamma_i\rangle$ could require an unbounded additional modulation variance $\sigma_{\gamma_i}^2$.

To avoid this problem, the precoding operator $V$ defines the $\Omega$ equivalence class set of the $|\varphi_i\rangle$ Gaussian CV states in the phase space with $d_{EC}$ equivalence class $EC_k|\varphi_i\rangle$ CV states. Denoting the CV states of a set $\Omega$ by $EC_k|\varphi_i\rangle$, $\Omega$ is defined as

$$\Omega \equiv \left\{EC_k|\varphi_i\rangle\right\}_{k=1}^{d_{EC}}. \tag{111}$$

The phase space constellation $\mathcal{C}_\mathcal{S}$ constructed by $V$ is the superset of $\Omega$ equivalence class sets, as follows:

$$\mathcal{C}_\mathcal{S} \equiv \bigcup_D \Omega = \bigcup_D \left\{EC_k|\varphi_i\rangle\right\}_{k=1}^{d_{EC}}, \tag{112}$$

where $D$ refers to the phase space domain where $|\kappa_i\rangle$ (more precisely, $U_2^{-1}|\gamma_i\rangle$) could occur in $\mathcal{S}$.

The naive approach is illustrated in Fig. 5.

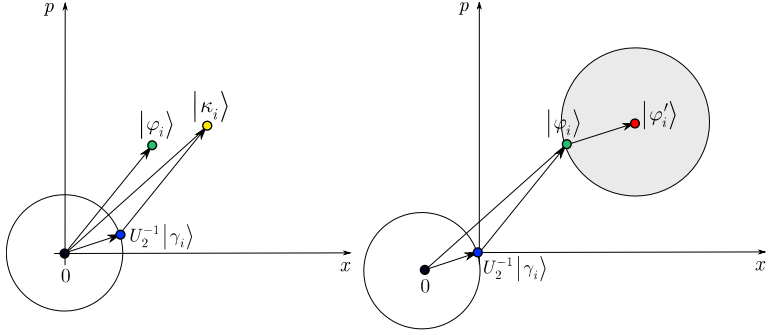

**Figure 5.** The naive precoding approach. (a) The transmitted $|\kappa_i\rangle$ is the difference of the user Gaussian CV $|\varphi_i\rangle$ and the interference $U_2^{-1}|\gamma_i\rangle$. The naive approach is strictly suboptimal because the energy increase $\sigma_{\gamma_i}^2$ required for $|\kappa_i\rangle = |\varphi_i\rangle - U_2^{-1}|\gamma_i\rangle$ could be unbounded, depending on the difference of $|\varphi_i\rangle$ and $U_2^{-1}|\gamma_i\rangle$. (b) The CV state $|\kappa_i\rangle$ under interference $U_2^{-1}|\gamma_i\rangle$ is transformed into $|\varphi_i\rangle$. The origin at the decoder is shifted onto $U_2^{-1}|\gamma_i\rangle$.

Operator $V$ chooses an $EC_k|\varphi_i\rangle$ from $\Omega$, for which $EC_k|\varphi_i\rangle - U_2^{-1}|\gamma_i\rangle$ is minimal, and prepares a transmit Gaussian CV state $|\kappa_i\rangle$ from $EC_k|\varphi_i\rangle$ such that
$$|\kappa_i\rangle = \min\left\{EC_k|\varphi_i\rangle - U_2^{-1}|\gamma_i\rangle\right\}. \tag{113}$$
In particular, in this approach, the required modulation variance $\sigma_{\gamma_i}^2$ is bounded because although the difference $|\varphi_i\rangle - U_2^{-1}|\gamma_i\rangle$ can be arbitrary, by choosing the closest $EC_k|\varphi_i\rangle$ to $U_2^{-1}|\gamma_i\rangle$, the difference of $EC_k|\varphi_i\rangle - U_2^{-1}|\gamma_i\rangle$ is restricted and minimized, which also results in a minimized additional modulation variance $\sigma_{\gamma_i}^2$.

We define an optimal phase space constellation $\mathcal{C}_\mathcal{S}$ for $V$, which leads to capacity-achieving communication over the over the Gaussian quantum channel $\mathcal{N}$. We formulate the problem directly in terms of $|\varphi_i\rangle$ single-carrier Gaussian CVs. The proof assumes a modulation variance $\sigma_\omega^2$ for the $|\varphi_i\rangle$ transmit Gaussian CVs, with total constraint $\frac{1}{l}\sum_l \sigma_{\omega_i}^2 = \sigma_\omega^2$.

By a simple sphere packing argument, for a $d$-dimensional random phase space code $\varsigma = \{|\varphi_i\rangle\}_{i=1}^d$, at a modulation variance $\sigma_\omega^2$, all of these Gaussian CV states lie in a sphere $\Theta$ of radius $\sqrt{d\sigma_\omega^2}$. From the law of large numbers, the received noisy phase space code $\varsigma' = \{|\varphi_i'\rangle\}_{i=1}^d$ will lie within a sphere $\Theta$ of radius $\sqrt{d\left(\sigma_\omega^2 + \sigma_\mathcal{N}^2\right)}$, and a noisy $|\varphi_i'\rangle$ will lie in a sphere $\Theta$ of radius $\sqrt{\sigma_\mathcal{N}^2}$ around the transmitted CV state $|\varphi_i\rangle$.

Assuming a maximum likelihood nearest neighbor decision [18-20] rule decoder $\mathcal{D}$ (referred to as $\mathcal{D}_\alpha$ throughout), $|\varphi_i'\rangle$ will be decoded to the transmit Gaussian CV state $|\varphi_i\rangle$ closest to $\alpha|\varphi_i'\rangle$, where

$$\alpha = \frac{\sigma_\omega^2}{\sigma_\omega^2 + \sigma_\mathcal{N}^2}, \tag{114}$$

which is then, for a $d$-dimensional case precisely leads to

$$\begin{aligned} \left\| \alpha \varsigma' - \varsigma \right\|^2 &= \left\| \alpha F(\Delta) + (\alpha - 1)\varsigma \right\|^2 \\ &= \alpha^2 d\sigma_\mathcal{N}^2 + (\alpha - 1) d\sigma_\omega^2 \\ &= \frac{d \sigma_\omega^2 \sigma_\mathcal{N}^2}{\sigma_\omega^2 + \sigma_\mathcal{N}^2}. \end{aligned} \tag{115}$$

In particular, for decoder $\mathcal{D}_\alpha$, the transmit Gaussian $|\varphi_i\rangle = EC_k |\varphi_i\rangle$ lies inside an $\Theta$ around $\alpha |\varphi_i'\rangle$ of the radius,

$$r = \sqrt{\frac{\sigma_\omega^2 \sigma_\mathcal{N}^2}{\sigma_\omega^2 + \sigma_\mathcal{N}^2}}. \tag{116}$$

Specifically, the probability $p$ that another random Gaussian CV $|\varphi_j\rangle$, $j \neq i$, lies inside the $\Theta$ of $\alpha |\varphi_i'\rangle$ is as follows:

$$p = \sqrt{\frac{\sigma_\omega^2 \sigma_\mathcal{N}^2}{\sigma_\omega^2 + \sigma_\mathcal{N}^2}} \frac{1}{\sqrt{\sigma_\omega^2}} = \sqrt{\frac{\sigma_\mathcal{N}^2}{\sigma_\omega^2 + \sigma_\mathcal{N}^2}}. \tag{117}$$

Thus, for $\alpha \varsigma'$, this probability is

$$p = \left( \frac{\sigma_\mathcal{N}^2}{\sigma_\omega^2 + \sigma_\mathcal{N}^2} \right)^{d/2}. \tag{118}$$

By the union bound, for $K_\varsigma$ $d$-dimensional $\varsigma_k$-s, $\{\varsigma_k\}_{k=1}^{K_\varsigma}$,

$$K_\varsigma < \frac{1}{p} = \left( \frac{\sigma_\mathcal{N}^2}{\sigma_\omega^2 + \sigma_\mathcal{N}^2} \right)^{-d/2}, \tag{119}$$

the achievable maximal rate in AMQD is calculated as follows:

$$C_{AMQD}(\mathcal{N}) = \frac{\log_2 K_\varsigma}{d} = \frac{\log_2 \frac{1}{p}}{d} - \frac{\log_2 d}{d} = \mathbb{E}\left[ \sum_l \frac{1}{2} \log_2 \left( 1 + \frac{\sigma_{\omega_i}^2 |F(T_i(\mathcal{N}_i))|^2}{\sigma_\mathcal{N}^2} \right) \right]. \tag{120}$$

After some calculations, by choosing $\alpha$ as in Equation (114), the decoding error probability [18] of $\mathcal{D}_\alpha$ is

$$p_{err}^{\mathcal{D}_\alpha} = \frac{\sigma_\omega^2 \sigma_\mathcal{N}^2}{\sigma_\omega^2 + \sigma_\mathcal{N}^2}, \tag{121}$$

which provably achieves the capacity of the Gaussian quantum channel $\mathcal{N}$.

The appropriate constellation $\mathcal{C}_\mathcal{S}$ covers uniformly the entire $\mathcal{S}$ containing the replicas of symbols $|\varphi_i\rangle$ in the full domain $\mathrm{D}$. The relative distance between $|\varphi_i\rangle$-s is preserved in each $\Omega$. At this point, the naive precoding can be revised.

The received noisy Gaussian CV $\alpha |\varphi_i'\rangle$ can be rewritten as

$$\begin{aligned} \alpha |\varphi_i'\rangle &= \alpha \left( |\varphi_1\rangle + U_2^{-1} |\gamma_i\rangle + U |\Delta_i\rangle \right) \\ &= \alpha \left( |\varphi_1\rangle + U |\Delta_i\rangle \right) + \alpha U_2^{-1} |\gamma_i\rangle, \end{aligned} \tag{122}$$

Hence, the transmit Gaussian CV from Equation (113) is reevaluated as

$$|\tilde{\kappa}_i\rangle = \min \left\{ EC_k |\varphi_i\rangle - \alpha U_2^{-1} |\gamma_i\rangle \right\}, \tag{123}$$

where
$$EC_k |\varphi_i\rangle = |\tilde{\kappa}_i\rangle + \alpha U_2^{-1} |\gamma_i\rangle \quad (124)$$
and
$$\alpha |\varphi_i'\rangle = EC_k' |\varphi_i\rangle, \quad (125)$$
such that
$$EC_k |\varphi_i\rangle - EC_k' |\varphi_i\rangle = |\varphi_i\rangle - \alpha(|\varphi_i\rangle + U|\Delta_i\rangle) \\ = |\varphi_i\rangle - \alpha |\varphi_i'\rangle. \quad (126)$$

The result in Equation (123) means that in the precoding process, the encoder pre-compensates $EC_k |\varphi_i\rangle$ by $-\alpha U_2^{-1} |\gamma_i\rangle$, instead of $-U_2^{-1} |\gamma_i\rangle$, for the given CV state $EC_k |\varphi_i\rangle$ of $\mathcal{C}_\mathcal{S}$ closest to $\alpha F |\gamma_i\rangle$.

An equivalence class set $\Omega$ (see Equation (111)) contains the replicas of the $|\varphi_i\rangle$ phase space symbols such that the $EC_k |\varphi_i\rangle$, $i = 1,...,|\Omega| d_{EC}$ of $\mathcal{C}_\mathcal{S}$ cover the entire phase space domain $D$, where $\alpha U_2^{-1} |\gamma_i\rangle$ could occur. Without loss of generality, the modulation variance increase $\sigma_{\gamma_i}^2$ of Equation (123) within $D$ is upper bounded as follows:
$$|\tilde{\kappa}_i\rangle = \max \min \left\{ EC_k |\varphi_i\rangle - \alpha U_2^{-1} |\gamma_i\rangle \right\}. \quad (127)$$

The decoding process of $\mathcal{D}_\alpha$ decides on $EC_k |\varphi_i\rangle = |\varphi_i\rangle$ closest to $\alpha |\varphi_i'\rangle$. Note that in the naive approach, the a decoder $\mathcal{D}$ with no constraint of (114) decides on $|\varphi_i'\rangle$ instead of $\alpha |\varphi_i'\rangle$, thus for $\mathcal{D}$ the uncertainty sphere $\Theta$ around $|\varphi_i'\rangle$ will lie outside the $\sigma_\omega$-radius sphere $\Theta$ of $|\varphi_i\rangle$, resulting in the radius,
$$\tilde{r} = \sigma_\omega + \sigma_\mathcal{N} > \sigma_\omega. \quad (128)$$

In particular, by a random coding argumentation, the $h_{\min}$ minimal distance between $|\varphi_i\rangle$ and $|\varphi_j\rangle, j \neq i$, is $h_{\min} = 2\sigma_\omega$ because at $\sigma_\omega^2$, a random CV $|\varphi_i\rangle$ lies in an uncertainty sphere of radius $\sigma_\omega$. For $\mathcal{D}_\alpha$, the radius of $\Theta$ around $\alpha |\varphi_i'\rangle$ is $r$, see Equation (116), thus $\Theta$ lives inside the $\sigma_\omega$-radius sphere.

The precoding operator $V$ includes the construction of an optimal $\mathcal{C}_\mathcal{S}$ and the encoding of $|\tilde{\kappa}_i\rangle = \min \left\{ EC_k |\varphi_i\rangle - \alpha U_2^{-1} |\gamma_i\rangle \right\}$, where $\alpha \to 0$ in the low-SNR regime; thus,
$$|\tilde{\kappa}_i\rangle = EC_k |\varphi_i\rangle. \quad (129)$$

From the structure of $\Omega$, the optimality of the precoding operator $V$ straightforwardly follows because the $EC_k |\varphi_i\rangle$ phase space symbols in $\mathcal{S}$ are restricted to domain $D$.

In the first setting of AMQD-MQA, operation $V$ is performed on the $d_i$ variables that identify the $|\phi_i\rangle$ CV states in the phase space. It is illustrated in Fig. 6.

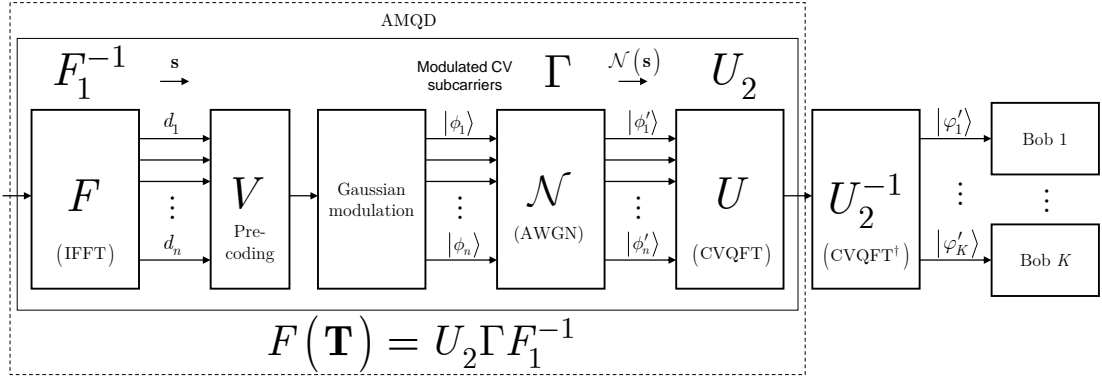

**Figure 6.** The first setting of AMQD-MQA with a precoding operator $V$. The SVD of AMQD is expressed by the matrix $F(\mathbf{T}) = U_2 \Gamma F_1^{-1}$.

As depicted in Fig. 7, in the second setting of AMQD-MQA, the users transmit single-carrier Gaussian CV states to the encoder $\mathcal{E}$. Hence, operation $V$ acts on the $EC_k |\varphi_i\rangle = |\varphi_i\rangle$ Gaussian CV states. After the precoding process, the CV subcarriers are sent through the $\mathcal{N}_i$ Gaussian sub-channels, see Section 3.

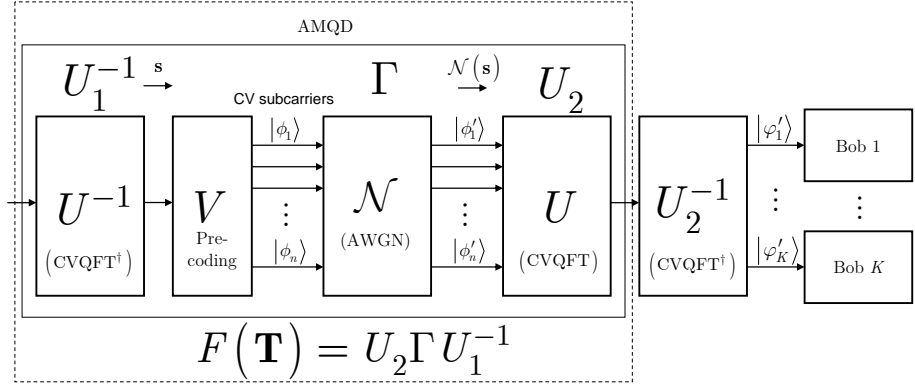

**Figure 7.** The second setting of AMQD-MQA with precoding operator $V$. The SVD of AMQD is expressed by the matrix $F(\mathbf{T}) = U_2 \Gamma U_1^{-1}$.

For $\mathcal{D}_\alpha$, the number of uncertainty spheres that can be packed into the sphere of radius $\sqrt{\sigma_\omega^2}$, by a simple sphere packing argument, yields

$$C_{AMQD}(\mathcal{N}) = \log_2 \frac{Vol\left(\Theta\left(\sqrt{\sigma_\omega^2}\right)\right)}{Vol\left(\Theta\left(\sqrt{\frac{\sigma_\omega^2 \sigma_\mathcal{N}^2}{\sigma_\omega^2 + \sigma_\mathcal{N}^2}}\right)\right)}, \quad (130)$$

where $Vol(\cdot)$ stands for the volume of the sphere $\Theta$. As it can be concluded, within the AMQD framework, the formula of Equation (130) exactly coincidences with the capacity of the $l$ Gaussian sub-channels $\mathcal{N}_i$ [4].

The phase space constellation $\mathcal{C}_\mathcal{S}$ is depicted in Fig. 8. For the $\mathcal{D}_\alpha$ decoder, the $r$-radius uncertainty sphere $\Theta$ around $\alpha |\varphi_i'\rangle$ will lie in the $\sigma_\omega$-radius sphere around $EC_k |\varphi_i\rangle$.

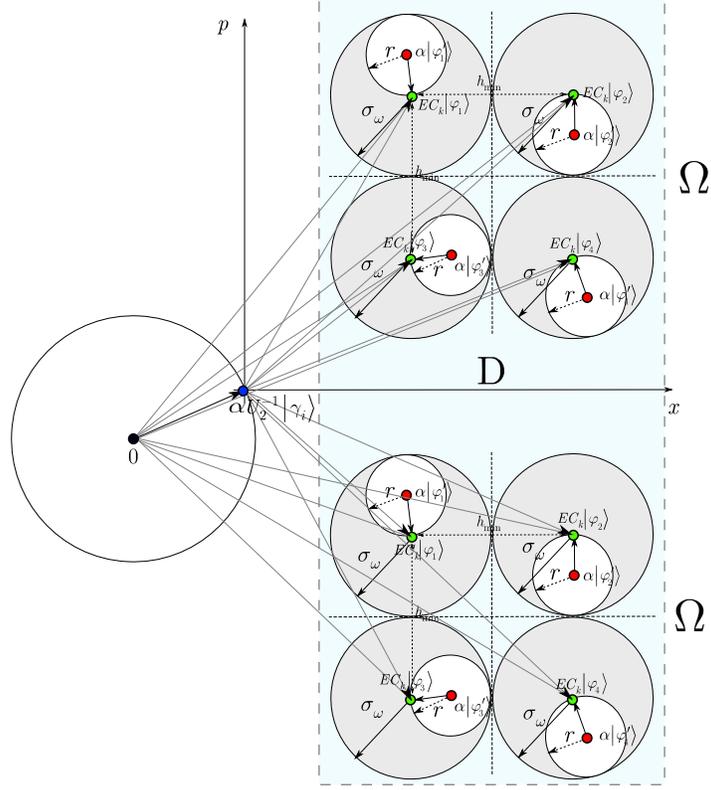

**Figure 8.** The phase space constellation $\mathcal{C}_{\mathcal{S}} = \bigcup_D \Omega = \bigcup_D \{EC_k |\varphi_i\rangle\}_{k=1}^{d_{EC}}$ at the use of the precoding operator $V$. The CV states of $\Omega$ are labeled as $EC_k |\varphi_i\rangle$. The $h_{\min}$ minimal distance between the constellation points is $2\sigma_\omega$. In the encoding phase, the precoding is performed via the Fourier-transformed interference $\alpha U_2^{-1} |\gamma_i\rangle$, where $\alpha = \sigma_\omega^2 / \sigma_\omega^2 + \sigma_{\mathcal{N}}^2$, and a random $|\varphi_i\rangle$ lies in a sphere of radius $\sigma_\omega$. The eigenchannel interference shifts the origin into $\alpha U_2^{-1} |\gamma_i\rangle$ at $\mathcal{D}_\alpha$. In decoding, the transmit Gaussian CV $|\varphi_i\rangle$ lies in the uncertainty sphere of radius $r = \sqrt{\sigma_\omega^2 \sigma_{\mathcal{N}}^2 / \sigma_\omega^2 + \sigma_{\mathcal{N}}^2}$ around $\alpha |\varphi_i'\rangle$. The noisy $\alpha |\varphi_i'\rangle$ will be decoded to the transmit CV state $|\varphi_i\rangle$ closest to $\alpha |\varphi_i'\rangle$ ($x$, position quadrature; $p$, momentum quadrature).

Note that for constellation $\mathcal{C}_{\mathcal{S}}$, there is no another $EC_k |\varphi_j\rangle, j \neq i$ constellation point within the uncertainty sphere of $\alpha |\varphi_i'\rangle$. Using a decoder $\mathcal{D}$ (with no $\alpha$-constraint), the received noisy $|\varphi_i'\rangle$ would lie outside the $\sigma_\omega$ sphere of $|\varphi_i\rangle$.

Let us to focus specifically on the crucial low-SNR regime. By taking the limit
$$\lim_{\sigma_{\mathcal{N}}^2 \to \infty} \alpha \to 0, \tag{131}$$

Equation (123) picks up the formula of
$$|\tilde{\kappa}_i\rangle = \min\{EC_k |\varphi_i\rangle\}. \tag{132}$$

Thus,

$$\alpha U_2^{-1} |\gamma_i\rangle = \varnothing \tag{133}$$

and

$$|\tilde{\kappa}_i\rangle = EC_k |\varphi_i\rangle. \tag{134}$$

The a reliable transmission over an Gaussian channel strictly lower bounds the SNR, particularly SNR $\geq$ -1.59 dB [18-20]. In fact, Equations (131)–(134) are also lower bounded by this constraint.

To summarize, at $V$, the achievable overall capacity in the singular layer at a partial channel side information $S(F(\mathbf{T}))$, with a covariance matrix

$$\mathbf{K}_\mathbf{s} = Q diag\left\{\sigma^2_{\omega'_1}, \ldots, \sigma^2_{\omega'_{K_{in}}}\right\} Q^\dagger \tag{135}$$

is as follows:

$$C(\mathcal{N}) = \mathbb{E}\left[\log_2 \det\left(\mathbf{I}_{K_{out}} + \tfrac{1}{\sigma^2_\mathcal{N}} F(\mathbf{T}) \mathbf{K}_\mathbf{s} F(\mathbf{T})^\dagger\right)\right], \tag{136}$$

thus, the available transmission rate is

$$R(\mathcal{N}) \leq \mathbb{E}\left[\log_2 \det\left(\mathbf{I}_{K_{out}} + \tfrac{1}{\sigma^2_\mathcal{N}} F(\mathbf{T}) \mathbf{K}_\mathbf{s} F(\mathbf{T})^\dagger\right)\right]. \tag{137}$$

∎

## 4.4 Phase space decoding of Gaussian random quadratures

**Lemma 3**. *The decoder $\mathcal{D}_\alpha$ is optimal for the phase space decoding of Gaussian random quadratures since it maximizes the SNIR.*

*Proof.*

Let assume that user $U_k$ utilizes an $\mathcal{D}_\alpha$ decoder for his $d$-dimensional output $\mathbf{y}_k = [y_{k,1}, \ldots, y_{k,d}]^T$. Let $\mathcal{N}^{U_k}$ the $d$-dimensional channel vector of user $U_k$, and $\chi_k$ to be the sum of noise $U(\Delta_k)$ and the $U_2^{-1}(\gamma_k)$ interference terms

$$\chi_k = U(\gamma_k) + U_2^{-1}(\Delta_k), \tag{138}$$

with invertible covariance matrix

$$\mathbf{K}_{\chi_k} = \left(\sigma^2_\mathcal{N} + \sigma^2_\gamma\right) \mathbf{I}_{K_{out}} + \sum_{j \neq k}^{n_{\min}} \sigma^2_{\omega'_j} F\left(\mathbf{T}_j\left(\mathcal{N}^{U_j}\right)\right) F\left(\mathbf{T}_j\left(\mathcal{N}^{U_j}\right)\right)^\dagger, \tag{139}$$

for which

$$\left(\mathbf{K}_{\chi_k} + \mathbf{x}\mathbf{x}^\dagger\right)^{-1} = \mathbf{K}_{\chi_k}^{-1} - \frac{\mathbf{K}_{\chi_k}^{-1} \mathbf{x}\mathbf{x}^\dagger \mathbf{K}_{\chi_k}^{-1}}{1 + \mathbf{x}^\dagger \mathbf{K}_{\chi_k}^{-1} \mathbf{x}}, \tag{140}$$

which further can be decomposed as

$$\mathbf{K}_{\chi_k} = U \Sigma U^\dagger, \tag{141}$$

where $U$ is a unitary, and $\Sigma$ is a diagonal matrix.

User $U_k$'s $\mathcal{D}_\alpha$ decoder first utilizes

$$\mathbf{K}_{\chi_k}^{0.5} = U \sum\nolimits^{0.5} U^{\dagger}, \tag{142}$$

where $\sum^{0.5}$ is a diagonal matrix with the square rooted elements of $\sum$, and $\mathbf{K}_{\chi_k}^{-0.5} F\left(\mathbf{T}_k\left(\mathcal{N}^{U_k}\right)\right)$ on his system $\mathbf{y}_k = F\left(\mathbf{T}_k\left(\mathcal{N}^{U_k}\right)\right) z_k + U_2^{-1}\left(\gamma_k\right) + U\left(\Delta_k\right)$, where $U_2^{-1}\left(\gamma_k\right) \in \mathcal{CN}\left(0, \mathbf{K}_{\gamma_k}\right)$, and $F\left(\Delta_k\right) \in \mathcal{CN}\left(0, \mathbf{K}_{\Delta_k}\right)$, which precisely leads to

$$\begin{aligned}
&\left(\mathbf{K}_{\chi_k}^{-0.5} F\left(\mathbf{T}_k\left(\mathcal{N}^{U_k}\right)\right)\right)^{\dagger} \mathbf{K}_{\chi_k}^{-0.5} \mathbf{y}_k \\
&= \left(\mathbf{K}_{\chi_k}^{-0.5} F\left(\mathbf{T}_k\left(\mathcal{N}^{U_k}\right)\right)\right)^{\dagger} \mathbf{K}_{\chi_k}^{-0.5} F\left(\mathbf{T}_k\left(\mathcal{N}^{U_k}\right)\right) z_k + \left(\mathbf{K}_{\chi_k}^{-0.5} F\left(\mathbf{T}_k\left(\mathcal{N}^{U_k}\right)\right)\right)^{\dagger} \mathbf{K}_{\chi_k}^{-0.5} \chi_k \\
&= F\left(\mathbf{T}_k\left(\mathcal{N}^{U_k}\right)\right)^{\dagger} \mathbf{K}_{\chi_k}^{-1} F\left(\mathbf{T}_k\left(\mathcal{N}^{U_k}\right)\right) z_k + F\left(\mathbf{T}_k\left(\mathcal{N}^{U_k}\right)\right)^{\dagger} \mathbf{K}_{\chi_k}^{-1} \left(U\left(\Delta_k\right) + U_2^{-1}\left(\gamma_k\right)\right),
\end{aligned} \tag{143}$$

where $z_k \in \mathcal{CN}\left(0, 2\sigma_{\omega_0}^2\right)$, thus $\mathcal{D}_{\alpha}$ can be rewritten as

$$\mathcal{D}_{\alpha} = \mathbf{K}_{\chi_k}^{-1} F\left(\mathbf{T}_k\left(\mathcal{N}^{U_k}\right)\right). \tag{144}$$

In particular, the result in Equation (144) satisfies that

$$I\left(z_k : \mathbf{y}_k\right) = I\left(z_k : \mathcal{D}_{\alpha}^{\dagger} \mathbf{y}_k\right), \tag{145}$$

where $z_k$ and $\mathbf{y}_k$ are independent conditioned on $\mathcal{D}_{\alpha}^{\dagger} \mathbf{y}_k$.

In fact, in Equation (145), $\mathcal{D}_{\alpha}^{\dagger} \mathbf{y}_k$ provides a sufficient statistic $S\left(\mathcal{D}_{\alpha}^{\dagger} \mathbf{y}_k\right)$ to decode $z_k$ from $\mathbf{y}_k$, since projecting the vector $\mathbf{y}_k$ onto the unit vector

$$\mathbf{v}_k = F\left(\mathbf{T}_k\left(\mathcal{N}^{U_k}\right)\right) \Big/ \left\| F\left(\mathbf{T}_k\left(\mathcal{N}^{U_k}\right)\right) \right\| \tag{146}$$

yields

$$\mathbf{v}_k^{\dagger} \mathbf{y}_k = \left\| F\left(\mathbf{T}_i\left(\mathcal{N}^{U_k}\right)\right) \right\| z_k + \chi_k, \tag{147}$$

where $\chi_i \in \mathcal{CN}\left(0, 2\sigma_{\mathcal{N}}^2 + \sigma_{\gamma}^2\right)$.

The optimality also requires from $\mathcal{D}_{\alpha}$ that the estimate $\tilde{z}_k$ of $z_k$ from vector $\mathbf{y}_k$ minimizes the mean square error $\mathbb{E}\left[\left|z_k - \tilde{z}_k\right|^2\right]$.

The estimation $\tilde{z}_k$ can be rewritten as

$$\tilde{z}_k = A\mathbf{c}^{\dagger} \mathbf{y}_k = A\mathbf{c}^{\dagger} F\left(\mathbf{T}_k\left(\mathcal{N}^{U_k}\right)\right) z_k + A\mathbf{c}^{\dagger} \chi_k, \tag{148}$$

where $\|\mathbf{c}\| = 1$, where $A$ is a constant [18] expressed as

$$A = \frac{\mathbb{E}\left[\left|z_k\right|^2\right] \left|\mathbf{c}^{\dagger} F\left(\mathbf{T}_k\left(\mathcal{N}^{U_k}\right)\right)\right|^2}{\mathbb{E}\left[\left|z_k\right|^2\right] \left|\mathbf{c}^{\dagger} F\left(\mathbf{T}_k\left(\mathcal{N}^{U_k}\right)\right)\right|^2 + \mathbf{c}^{\dagger} \mathbf{K}_{\chi_k} \mathbf{c}} + \frac{F\left(\mathbf{T}_k\left(\mathcal{N}^{U_k}\right)\right)^{\dagger} \mathbf{c}}{\left|F\left(\mathbf{T}_k\left(\mathcal{N}^{U_k}\right)\right)^{\dagger} \mathbf{c}\right|}, \tag{149}$$

which leads to

$$\begin{aligned}
\frac{\mathbb{E}\left[\left|z_k\right|^2\right]}{\mathrm{E}} &= 1 + \frac{\mathbb{E}\left[\left|z_k\right|^2\right] \left|\mathbf{c}^{\dagger} F\left(\mathbf{T}_k\left(\mathcal{N}^{U_k}\right)\right)\right|^2}{\mathbf{c}^{\dagger} \mathbf{K}_{\chi_k} \mathbf{c}} \\
&= 1 + \frac{\mathbb{E}\left[\left|z_k\right|^2\right] \left|\mathbf{c}^{\dagger} F\left(\mathbf{T}_k\left(\mathcal{N}^{U_k}\right)\right)\right|^2}{\mathbf{c}^{\dagger} \left(\mathbf{K}_{\Delta} + \mathbf{K}_{\gamma_i}\right) \mathbf{c}}.
\end{aligned} \tag{150}$$

As follows, for

$$\mathbf{c} = \mathbf{K}_{\chi_k}^{-1} F\left(\mathbf{T}_k\left(\mathcal{N}^{U_k}\right)\right), \tag{151}$$

the error E is minimized as

$$\mathrm{E} = \frac{\mathbb{E}\left[|z_k|^2\right]}{1+\mathrm{SNIR}\left|F\left(\mathbf{T}_k\left(\mathcal{N}^{U_k}\right)\right)\right|^2}, \tag{152}$$

thus the SNIR is maximized in

$$\mathrm{SNIR} = \sigma_{\omega''}^2 \mathbf{K}_{\chi_k}^{-1}. \tag{153}$$

Particularly, in the low-SNR regimes

$$\sum_{j \neq k}^{n_{\min}} \sigma_{\omega_j'}^2 F\left(\mathbf{T}_j\left(\mathcal{N}^{U_j}\right)\right) F\left(\mathbf{T}_j\left(\mathcal{N}^{U_j}\right)\right)^{\dagger} \to 0, \tag{154}$$

thus Equation (139) can be re-evaluated as

$$\mathbf{K}_{\chi_k} = \sigma_{\mathcal{N}}^2 \mathbf{I}_{K_{out}}. \tag{155}$$

Specifically, from the covariance matrix $\mathbf{K}_{\chi_k}$, the signal-to-noise-plus-interference ratio in the multicarrier transmission is

$$\mathrm{SNIR} = \mathrm{SNR} = \mathbb{E}\left[\sigma_{\omega''}^2 \left(\sigma_{\mathcal{N}}^2 \mathbf{I}_{K_{out}}\right)^{-1}\right]. \tag{156}$$

Precisely, if $F\left(\mathbf{T}\left(\mathcal{N}^{U_k}\right)\right)$ is time-invariant then the total rate that can be achieved by $K_{out}$ instances of $\mathcal{D}_\alpha$ equals to the overall mutual information.

We apply the chain rule [18] of mutual information, from which

$$\begin{aligned} I\left(\mathbf{z}:\mathbf{y}\right) &= I\left(z_1,\ldots,z_{K_{in}}:\mathbf{y}\right) \\ &= I\left(z_1:\mathbf{y}\right) + I\left(z_2:\mathbf{y}\big|z_1\right) + \ldots + I\left(z_{K_{in}}:\mathbf{y}\big|z_1,\ldots,z_{K_{in}-1}\right) \end{aligned} \tag{157}$$

where

$$I\left(z_k:\mathbf{y}\big|z_1,\ldots,z_{k-1}\right) = I\left(z_k:\mathbf{y}'\right), \tag{158}$$

such that

$$\begin{aligned} \mathbf{y}' &= \mathbf{y} - \sum_{j=1}^{k-1} F\left(\mathbf{T}_j\left(\mathcal{N}^{U_j}\right)\right) z_j \\ &= F\left(\mathbf{T}_k\left(\mathcal{N}^{U_k}\right)\right) z_k + \sum_{j>k} F\left(\mathbf{T}_j\left(\mathcal{N}^{U_j}\right)\right) + U\left(\Delta\right). \end{aligned} \tag{159}$$

Thus, Equation (158) can be rewritten as

$$I\left(z_k:\mathbf{y}\big|z_1,\ldots,z_{k-1}\right) = I\left(z_k:\mathcal{D}_\alpha \mathbf{y}'\right), \tag{160}$$

where the resulting function $I\left(z_k:\mathbf{y}\big|z_1,\ldots,z_{k-1}\right)$ precisely picks up the information theoretic maximum.

∎

# 5 Phase Space Coding for the Gaussian Subcarrier CVs

**Theorem 3.** *The transmission of $|\phi_i\rangle$ Gaussian subcarrier CVs over the $\mathcal{N}_i$ Gaussian sub-channels can be optimized by a random phase space constellation $\mathcal{C}_\mathcal{S}(\mathcal{N})$ as $\mathcal{C}_\mathcal{S}^P(\mathcal{N}) = \left(\left|\phi_{1...d_{\mathcal{C}_\mathcal{S}^P(\mathcal{N}_1)}}\right\rangle, P_2\left|\phi_{1...d_{\mathcal{C}_\mathcal{S}^P(\mathcal{N}_2)}}\right\rangle, ..., P_l\left|\phi_{1...d_{\mathcal{C}_\mathcal{S}^P(\mathcal{N}_l)}}\right\rangle\right)$, where $P_i$, $i = 2,...,l$ is a random permutation operator, and $d_{\mathcal{C}_\mathcal{S}(\mathcal{N}_i)} = d_{\mathcal{C}_\mathcal{S}(\mathcal{N}_j)}$ is the cardinality of $\mathcal{C}_\mathcal{S}(\mathcal{N}_i)$. The optimality function of $\mathcal{C}_\mathcal{S}(\mathcal{N})$ is $o(\mathcal{C}_\mathcal{S}(\mathcal{N})) = \sum_l \left(\nu_{Eve} - |\delta_i|^2\right)$, where $\delta_i$ is the normalized codeword difference.*

*Proof.*
Assume that Alice and Bob agreed on the $l$ Gaussian sub-channels, $\mathcal{N}_i$. We rewrite again the output of $\mathcal{N}_i$ is as follows [3-4]:

$$y_i = F(T(\mathcal{N}_i))F(d_i) + U(\Delta_i), \tag{161}$$

where variable $d_i$ is the Gaussian subcarrier input of $\mathcal{N}_i$ (for further details, see Section 2.1 and [4]).

For two $l$-dimensional input code words $\mathbf{d}_A = (d_{A,1},...d_{A,l})^T$ and $\mathbf{d}_B = (d_{B,1},...d_{B,l})^T$, the probability that $\mathbf{d}_A$ is distorted onto $\mathbf{d}_B$ [18] on the output of the $l$ Gaussian sub-channels conditioned on $F(\mathbf{T}(\mathcal{N}))$ is as follows:

$$\Pr(\mathbf{d}_A \to \mathbf{d}_B | F(\mathbf{T}(\mathcal{N}))) = Q\left(\sqrt{\frac{\sigma_{\omega''}^2}{2\sigma_\mathcal{N}^2}\sum_l |F(T_i(\mathcal{N}_i))|^2 |\delta_i|^2}\right), \tag{162}$$

where $Q$ is the Gaussian tail function, $\sigma_{\omega''}^2$ is the modulation variance derived in Equation (75), and $\delta_i$ is the normalized difference [18] of $d_{A,i}$ and $d_{B,i}$, calculated as follows:

$$\delta_i = \frac{1}{\sqrt{\frac{\sigma_{\omega''}^2}{\sigma_\mathcal{N}^2}}}(d_{A,i} - d_{B,i}). \tag{163}$$

Assuming the case that in Equation (162), the condition

$$\frac{\sigma_{\omega''}^2}{2\sigma_\mathcal{N}^2}\sum_l |F(T_i(\mathcal{N}_i))|^2 |\delta_i|^2 < 1 \tag{164}$$

holds, one obtains

$$|\delta_{1...L}|^{2/l} > c\frac{1}{l2^R}. \tag{165}$$

Thus,

$$|\delta_{1...L}|^2 > c^l \frac{1}{l^l 2^{Rl}} \tag{166}$$

for any constant $c > 0$ and for an arbitrary pair of $\mathbf{d}_A$ and $\mathbf{d}_B$. Thus, Equation (165) is a sufficient condition on the optimality on the structure of the phase space constellation $\mathcal{C}_\mathcal{S}(\mathcal{N}_i)$.

In particular, at the achievable rate $R$ per $\mathcal{N}_i$, the cardinality of $\mathcal{C}_\mathcal{S}(\mathcal{N}_i)$ is as follows:

$$|\mathcal{C}_\mathcal{S}(\mathcal{N}_i)| = 2^R. \tag{167}$$

Thus, each $\mathcal{C}_\mathcal{S}(\mathcal{N}_i)$ is precisely defined with $2^R$ CV states $|\phi_i\rangle$ for each $\mathcal{N}_i$ Gaussian sub-channels, and $\mathcal{C}_\mathcal{S}(\mathcal{N}_i)$ contains $2^{lR}$ constellation points $|\phi_i\rangle$ in overall, that is,

$$|\mathcal{C}_\mathcal{S}(\mathcal{N})| = 2^{lR}. \tag{168}$$

Specifically, evaluating the $Q(\cdot)$ Gaussian tail function at $\min_{\forall F(T_i(\mathcal{N}_i))} F(T_i(\mathcal{N}_i))$ leads to the worst-case scenario as follows:

$$Q\left(\sqrt{\min_{\forall F(T_i(\mathcal{N}_i))} \frac{\sigma_{\omega''}^2}{2\sigma_\mathcal{N}^2} \sum_l |F(T_i(\mathcal{N}_i))|^2 |\delta_i|^2}\right), \tag{169}$$

such that for the $l$ $\mathcal{N}_i$ Gaussian sub-channels, the following constraint is satisfied:

$$\sum_l \log_2\left(1 + \frac{|F(\mathbf{T}(\mathcal{N}))|^2 \sigma_{\omega''}^2}{\sigma_\mathcal{N}^2}\right) = \sum_l \log_2\left(1 + \frac{|F(T_i(\mathcal{N}_i))|^2 \sigma_{\omega''}^2}{\sigma_\mathcal{N}^2}\right) \geq lR, \tag{170}$$

where $R$ is the average rate supported by each $\mathcal{N}_i$ Gaussian sub-channels.

Using Equation (163), the constraint of Equation (170) can be rewritten as follows [18]:

$$\sum_l \log_2\left(1 + \frac{\tilde{Q}_i(\mathcal{N}_i)}{|\delta_i|^2}\right) \geq lR, \tag{171}$$

where

$$\tilde{Q}_i(\mathcal{N}_i) = \frac{|F(T_i(\mathcal{N}_i))|^2 |\delta_i|^2 \sigma_{\omega''}^2}{\sigma_\mathcal{N}^2} \tag{172}$$

and

$$\min_{\forall \tilde{Q}_i \geq 0} \frac{1}{2} \sum_l \tilde{Q}_i = \min_{\forall F(T_i(\mathcal{N}_i))} \frac{\sigma_{\omega''}^2}{2\sigma_\mathcal{N}^2} \sum_l |F(T_i(\mathcal{N}_i))|^2 |\delta_i|^2. \tag{173}$$

In particular, from the modulation variance allocation algorithm of Section 3, some calculations reveal that $\min_{\forall F(T_i(\mathcal{N}_i))} \sum_l |F(T_i(\mathcal{N}_i))|^2$ can be reevaluated precisely as

$$\min_{\forall F(T_i(\mathcal{N}_i))} \sum_l |F(T_i(\mathcal{N}_i))|^2 = \sum_l \frac{1}{\sqrt{\frac{\sigma_{\omega''}^2}{\sigma_\mathcal{N}^2}}}\left(v_{Eve} \frac{1}{|\delta_i|^2} - 1\right). \tag{174}$$

From Equations (173) and (174), the Gaussian tail function in Equation (169) can be rewritten as [18]

$$Q\left(\sqrt{\frac{1}{2}\sum_l \left(v_{Eve} - |\delta_i|^2\right)}\right) \tag{175}$$

and

$$\sum_l \log_2\left(v_{Eve}\frac{1}{|\delta_i|^2}\right) = \sum_l \log_2\left(1 + \frac{\tilde{Q}_i(\mathcal{N}_i)}{|\delta_i|^2}\right) = lR. \tag{176}$$

This constraint on $v_{Eve}$ allows us to design phase space constellation for each $\mathcal{N}_i$ Gaussian sub-channels, such that the achievable rate $R$ over each sub-channel is maximized. With these results in mind, we design a phase space constellation for reliable simultaneous transmission over the Gaussian sub-channels, assuming the presence of an optimal Gaussian collective attack [4].

As one can readily conclude, the optimality of sub-channel phase space constellation $\mathcal{C}_{\mathcal{S}}(\mathcal{N})$ can be straightforwardly derived from Equations (175) and (176).

The optimality of $\mathcal{C}_{\mathcal{S}}(\mathcal{N})$ can explicitly be quantified via the *optimality function* $o(\cdot)$, calculated as follows:

$$o\big(\mathcal{C}_{\mathcal{S}}(\mathcal{N})\big) = \sum_{l}\Big(\nu_{Eve} - |\delta_i|^2\Big). \tag{177}$$

Assuming that $\mathcal{C}_{\mathcal{S}}(\mathcal{N})$ contains $l$ phase space constellations $\mathcal{C}_{\mathcal{S}}(\mathcal{N}_i)$ with $2^R$ $|\phi_i\rangle$ Gaussian subcarrier CVs in each $\mathcal{C}_{\mathcal{S}}(\mathcal{N}_i)$, but with independent $\mathcal{C}_{\mathcal{S}}(\mathcal{N}_i)$ sub-channel constellations for each $\mathcal{N}_i$, leads to [18]

$$\sum_{l-1}|\delta_i|^2 = \sum_{l-1}\left|\frac{1}{\sqrt{\frac{\sigma^2_{\omega''}}{\sigma^2_{\mathcal{N}}}}}(d_{A,i} - d_{B,i})\right|^2 = 0 \tag{178}$$

and

$$\sum_{l}\log_2\left(\nu_{Eve}\frac{1}{|\delta_i|^2}\right) = 0. \tag{179}$$

Thus, the optimality function converges to zero,

$$o\big(\mathcal{C}_{\mathcal{S}}(\mathcal{N})\big) = 0. \tag{180}$$

From Equation (180), it straightforwardly follows that independent phase space constellations $\mathcal{C}_{\mathcal{S}}(\mathcal{N}_i)$ for the Gaussian sub-channels $\mathcal{N}_i$ cannot lead to an optimal solution. In particular, a more sophisticated phase space constellation is required for a nonzero optimality function. Specifically, it can be achieved by the so-called random permutation coding [18-20], in which each $\mathcal{C}_{\mathcal{S}}(\mathcal{N}_i)$ of $\mathcal{N}_i$ maximizes $o\big(\mathcal{C}_{\mathcal{S}}(\mathcal{N}_i)\big)$ for $\forall i$.

In a phase space constellation $\mathcal{C}^P_{\mathcal{S}}$, each $\mathcal{N}_i$ Gaussian sub-channel has a different random phase space constellation $\mathcal{C}^P_{\mathcal{S}}(\mathcal{N}_i)$ such that if the two Gaussian subcarrier CVs $|\phi_i\rangle$ and $|\phi_j\rangle, j \neq i$ are close to each other in $\mathcal{C}^P_{\mathcal{S}}(\mathcal{N}_i)$ of $\mathcal{N}_i$, then in the $\mathcal{C}^P_{\mathcal{S}}(\mathcal{N}_j)$ of $\mathcal{N}_j$, the distance $\mho$ between $|\phi_i\rangle$ and $|\phi_j\rangle$ in $\mathcal{C}^P_{\mathcal{S}}(\mathcal{N}_j)$ is at least double the minimum distance of these Gaussian CVs in $\mathcal{C}^P_{\mathcal{S}}(\mathcal{N}_i)$.

Without loss of generality, for two sub-channels $\mathcal{N}_1$ and $\mathcal{N}_2$, with two reference constellation symbols $|\phi_i\rangle$ and $|\phi_j\rangle, j \neq i$ in $\mathcal{C}^P_{\mathcal{S}}(\mathcal{N}_1)$ of $\mathcal{N}_1$, the $\partial$ minimal distance function [18] is defined as follows:

$$\partial\big(\mathcal{C}^P_{\mathcal{S}}(\mathcal{N}_1)\big) = \min_{\forall d_{A,i}}(d_{A,i} - d_{A,j}), \tag{181}$$

whereas for sub-channel $\mathcal{N}_2$, the distance between $|\phi_i\rangle$ and $|\phi_j\rangle$ is precisely yielded as follows:

$$\partial\left(\mathcal{C}_{\mathcal{S}}^{P}\left(\mathcal{N}_{2}\right)\right) = \min_{\forall d_{A,i}}\left(d_{A,i} - d_{A,j}\right) \\ = \mho \cdot \partial\left(\mathcal{C}_{\mathcal{S}}^{P}\left(\mathcal{N}_{1}\right)\right), \tag{182}$$

where $\mho \geq 2$.

Smoothing the nonnegativity condition in Equation (173), a lower bound on $o\left(\mathcal{C}_{\mathcal{S}}^{P}\left(\mathcal{N}_{i}\right)\right)$ is yielded as follows [18]:

$$o\left(\mathcal{C}_{\mathcal{S}}^{P}\left(\mathcal{N}_{i}\right)\right) \geq l 2^{R}\left|\delta_{1\ldots l}\right|^{2/l} - \sum_{l}\left|\delta_{i}\right|^{2}, \tag{183}$$

where for $\nu'_{\min} = \sigma_{\mathcal{N}}^{2}\Big/\max_{n_{\min}} \lambda_{i}^{2} \geq 0$,

$$\sum_{l}\left|\delta_{i}\right|^{2} \to 0. \tag{184}$$

Thus, the lower bound is

$$o\left(\mathcal{C}_{\mathcal{S}}^{P}\left(\mathcal{N}_{i}\right)\right) \geq l 2^{R}\left|\delta_{1\ldots l}\right|^{2/l}. \tag{185}$$

The constellation $\mathcal{C}_{\mathcal{S}}^{P}\left(\mathcal{N}_{j}\right)$ of $\mathcal{N}_{j}$ is a result of a random one-to-one permutation map $P$ on $\mathcal{C}_{\mathcal{S}}^{P}\left(\mathcal{N}_{i}\right)$ of $\mathcal{N}_{i}, i \neq j$, as follows:

$$\mathcal{C}_{\mathcal{S}}^{P}\left(\mathcal{N}_{j}\right) = P\left(\mathcal{C}_{\mathcal{S}}^{P}\left(\mathcal{N}_{i}\right)\right). \tag{186}$$

For an $l$-dimensional input $\mathbf{d}$ of $\mathcal{N}_{1\ldots l}$, a random constellation point $\left|\phi_{i}\right\rangle$ in $\mathcal{C}_{\mathcal{S}}^{P}\left(\mathcal{N}_{1}\right)$ is selected for $\mathcal{N}_{1}$. Then for the remaining $l-1$ Gaussian sub-channel $\mathcal{N}_{2\ldots l}$, a random permutation map is applied for $P_{i}$ for $\mathcal{N}_{i}$.

Thus, the $d_{\mathcal{C}_{\mathcal{S}}(\mathcal{N}_{i})}$ constellation points $\left|\phi_{1\ldots d_{\mathcal{C}_{\mathcal{S}}(\mathcal{N}_{i})}}\right\rangle$ of $\mathcal{C}_{\mathcal{S}}^{P}\left(\mathcal{N}\right)$ formulate $\mathcal{C}_{\mathcal{S}}^{P}\left(\mathcal{N}\right)$ as follows:

$$\mathcal{C}_{\mathcal{S}}^{P}\left(\mathcal{N}\right) = \mathcal{C}_{\mathcal{S}}^{P}\left(\mathcal{N}_{1}\right),\ldots,\mathcal{C}_{\mathcal{S}}^{P}\left(\mathcal{N}_{l}\right) \\ = \left(\left|\phi_{1\ldots d_{\mathcal{C}_{\mathcal{S}}^{P}(\mathcal{N}_{1})}}\right\rangle, P_{2}\left|\phi_{1\ldots d_{\mathcal{C}_{\mathcal{S}}^{P}(\mathcal{N}_{2})}}\right\rangle, \ldots, P_{l}\left|\phi_{1\ldots d_{\mathcal{C}_{\mathcal{S}}^{P}(\mathcal{N}_{l})}}\right\rangle\right). \tag{187}$$

In particular, an important property of Equation (187) is that the $P_{i}$ permutation operators are drawn from a $\mathcal{U}$ uniform distribution, thus each $P_{i}$ operator occurs with probability

$$p = \tfrac{1}{2^{lR}}. \tag{188}$$

Some calculations yield that for the average inverse of Equation (166), the following upper bound holds for the $l-1$ random permutation operators $P_{i}$ taken on a pair of Gaussian subcarriers $\{d_{1}, d_{2}\}$:

$$\mathbb{E}_{P_{2,\ldots,l}}\left[\tfrac{1}{2^{lR}\left(2^{lR}-1\right)} \cdot \sum_{d_{1}\neq d_{2}} \tfrac{1}{\left|d_{1}-d_{2}\right|^{2}\left|P_{2}(d_{1})-P_{2}(d_{2})\right|^{2}\ldots\left|P_{l}(d_{1})-P_{l}(d_{2})\right|^{2}}\right] \leq l^{l}R^{l}, \tag{189}$$

where for the uniform at random $P_{i}, i = 2\ldots l$, permutation operators, the relation

$$\tfrac{1}{2^{lR}}\sum_{d_{1}}\left(\sum_{d_{1}\neq d_{2}} \tfrac{1}{\left|d_{1}-d_{2}\right|^{2}\left|P_{2}(d_{1})-P_{2}(d_{2})\right|^{2}\ldots\left|P_{l}(d_{1})-P_{l}(d_{2})\right|^{2}}\right) \leq l^{l}R^{l}2^{lR} \tag{190}$$

is straightforwardly yielded.

The structure of the phase space constellation $\mathcal{C}_\mathcal{S}^P(\mathcal{N}_i)$ for two sub-channels $\mathcal{N}_1$ and $\mathcal{N}_2$ with $d_{\mathcal{C}_\mathcal{S}^P(\mathcal{N}_i)} = 16$ [18] and $\mho = 2$, is illustrated in Fig. 9.

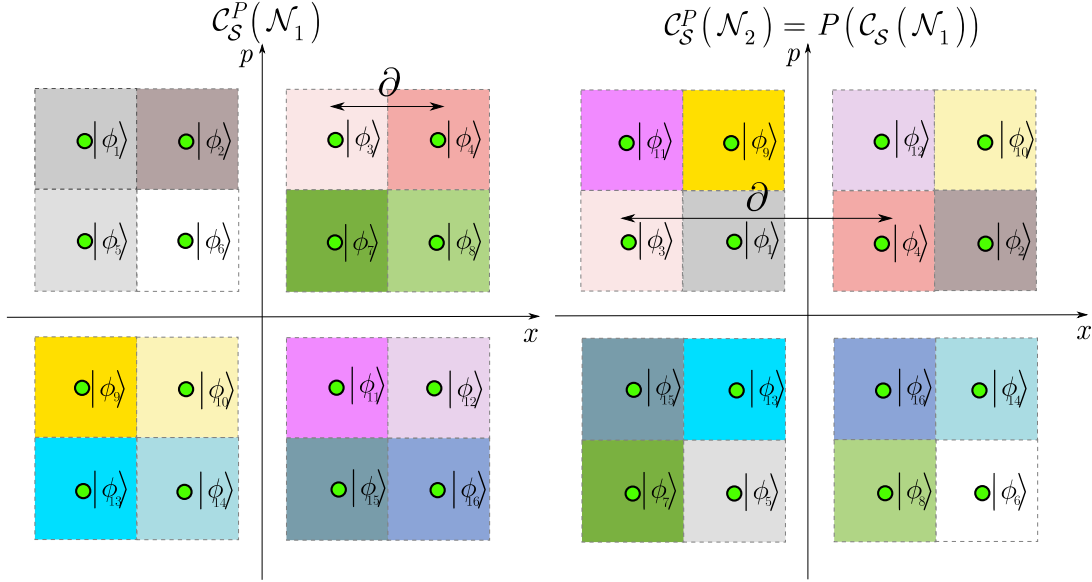

**Figure 9.** The phase space constellations $\mathcal{C}_\mathcal{S}^P(\mathcal{N}_1)$ and $\mathcal{C}_\mathcal{S}^P(\mathcal{N}_2)$ for Gaussian subcarrier CVs, at two Gaussian sub-channels $\mathcal{N}_1$ and $\mathcal{N}_2$, at $\mho = 2$. In each constellation, $d_{\mathcal{C}_\mathcal{S}^P(\mathcal{N}_i)} = 16$, $|\phi_i\rangle, i = 1,\ldots,d_{\mathcal{C}_\mathcal{S}(\mathcal{N}_i)}$, where $\mathcal{C}_\mathcal{S}^P(\mathcal{N}_2) = P(\mathcal{C}_\mathcal{S}^P(\mathcal{N}_1))$ and $\mathcal{C}_\mathcal{S}^P(\mathcal{N}_1) = P^{-1}\mathcal{C}_\mathcal{S}^P(\mathcal{N}_2)$, and $P$ is a random permutation operator drawn from a $\mathcal{U}$ uniform distribution ($x$, position quadrature; $p$, momentum quadrature).

Assuming that $|\delta_1| = \ldots = |\delta_l|$ is satisfied, the $Q$ function can be evaluated as follows:

$$Q\left(\sqrt{\tfrac{1}{2}\sum_l\left(\nu_{Eve} - |\delta_i|^2\right)}\right) = l\left(2^R - 1\right)|\delta_i|^2, \tag{191}$$

whereas for an ordered structure $|\delta_1| \leq \ldots \leq |\delta_l|$, with a largest index $j$ such that $|\delta_j|^{2j} \leq |\delta_{1\ldots j}|^2 \leq |\delta_{j+1}|^{2j}$, where $|\delta_{l+1}| = \infty$, Equation (173) is reduced to a maximization problem, as

$$\max\left(j\left(2^{Rl}|\delta_1\ldots\delta_j|^2\right)^{1/j} - \sum_{i=1}^{j}|\delta_i|^2\right). \tag{192}$$

Note that $j = l$ results in $|\delta_1| = \ldots = |\delta_l|$. Thus, the condition coincidences with Equation (191).

∎

# 6 Conclusions

We proposed a singular layer transmission model for CVQKD. The CVQKD protocols have been introduced for practical reasons, most importantly for their easy implementation over standard telecommunication networks and flexible encoding and decoding capabilities. The singular layer of CVQKD injects an extra degree of freedom into the transmission to extend the achievable distances and to improve the secret key rates. The benefits of the singular layer have been demonstrated through the AMQD and AMQD-MQA schemes, which use Gaussian subcarrier CVs for the transmission. We derived the rate formulas and studied the properties of an optimal phase space constellation. The SIA scheme provides a solution for the eigenchannel interference cancellation, which phenomenon occurs at a partial channel side information. We also analyzed the properties of an information-theoretical optimal phase space decoding and defined a phase space constellation for the Gaussian sub-channels, which is provably optimal and offers unconditional security for the transmission. The results confirm that the additional degree of freedom brought in by the singular layer can be significantly exploited in the crucial low-SNR regimes, which is particularly convenient for long-distance CVQKD scenarios.

# Acknowledgements

The author would like to thank Professor Sandor Imre for useful discussions. The results are supported by the grant COST Action MP1006.# References

[1]  S. Pirandola, S. Mancini, S. Lloyd, and S. L. Braunstein, Continuous-variable Quantum Cryptography using Two-Way Quantum Communication, *arXiv:quant-ph/0611167v3* (2008).

[2]  F. Grosshans, P. Grangier, Reverse reconciliation protocols for quantum cryptography with continuous variables, *arXiv:quant-ph/0204127v1* (2002).

[3]  L. Gyongyosi, Multiuser Quadrature Allocation for Continuous-Variable Quantum Key Distribution, *arXiv:1312.3614* (2013).

[4]  L. Gyongyosi, Adaptive Multicarrier Quadrature Division Modulation for Continuous-variable Quantum Key Distribution, *arXiv:1310.1608* (2013).

[5]  S. Pirandola, R. Garcia-Patron, S. L. Braunstein and S. Lloyd. *Phys. Rev. Lett.* 102 050503. (2009).

[6]  S. Pirandola, A. Serafini and S. Lloyd. *Phys. Rev. A* 79 052327. (2009).

[7]  S. Pirandola, S. L. Braunstein and S. Lloyd. *Phys. Rev. Lett.* 101 200504 (2008).

[8]  C. Weedbrook, S. Pirandola, S. Lloyd and T. Ralph. *Phys. Rev. Lett.* 105 110501 (2010).

[9]  C. Weedbrook, S. Pirandola, R. Garcia-Patron, N. J. Cerf, T. Ralph, J. Shapiro, and S. Lloyd. *Rev. Mod. Phys.* 84, 621 (2012).38

# Supplemental Information

## S.1 Notations

The notations of the manuscript are summarized in Table S.1.

**Table S.1.** Summary of notations.

| Notation | Description |
|---|---|
| $\mathcal{B}(\mathbf{M}) = U\Gamma V^{-1}$ | Bound operator for the extension of SVD onto a Hilbert space $\mathcal{H}$, $\mathcal{B}(\mathbf{M}) = UV^{-1} \cdot V\Gamma V^{-1}$, where $U$ is a partial isometry, $V$ is unitary, and a measure space $Z$. |
| $S_1$, $S_2$ | Sets of singular operators $S_1 = \{F_1, U_2^{-1}\}$, $S_2 = \{U_1, U_2^{-1}\}$. |
| $F(\mathbf{T}) = U_2 \Gamma F_1^{-1}$, $F(\mathbf{T}) = U_2 \Gamma U_1^{-1}$ | The SVD of $F(\mathbf{T})$, where $F_1^{-1}, F_1 \in \mathbb{C}^{K_{in} \times K_{in}}$ and $U_2, U_2^{-1} \in \mathbb{C}^{K_{out} \times K_{out}}$ are unitary matrices, $K_{in}$ and $K_{out}$ refer to the number of sender and receiver users such that $K_{in} \leq K_{out}$, $F_1^{-1}F_1 = F_1 F_1^{-1} = I$, $U_2 U_2^{-1} = U_2^{-1} U_2 = I$, and $\Gamma \in \mathbb{R}$ is a diagonal matrix with non-negative real diagonal elements $\lambda_i$, $F(\mathbf{T}) = \sum_{n_{\min}} \lambda_i U_{2,i} F_{1,i}^{-1}$. |
| $\lambda_1 \geq \lambda_2 \geq \ldots \lambda_{n_{\min}}$ | The non-negative real diagonal elements of the diagonal matrix $\Gamma \in \mathbb{R}$, called the eigenchannels of $F(\mathbf{T}) = U_2 \Gamma F_1^{-1}$. |
| $\lambda_i^2$ | The $n_{\min}$ squared eigenchannels $\lambda_i^2$ are the eigenvalues of $F(\mathbf{T})F(\mathbf{T})^\dagger = U_2 \Gamma \Gamma^T U_2^{-1}$. |
| $n_{\min}$ | $n_{\min} = \min(K_{in}, K_{out})$, equals to the rank of $F(\mathbf{T})$, where $K_{in} \leq K_{out}$. |
| $\mathbf{s}$ | Stream matrix, $\mathbf{s} = (s_1, \ldots, s_{n_{\min}})^T \in \mathcal{CN}(0, \mathbf{K_s})$, defined by the unitary $F_1$ ($U_1$) applied on $\mathbf{z} \in \mathcal{CN}(0, \mathbf{K_z})$. |
| $s_i$ | A stream variable $s_i$ that identifies the CV state $|s_i\rangle$ in the phase space $\mathcal{S}$. Expressed as $|s_i'\rangle = \lambda_i U_{2,i} F_{1,i}^{-1} |s_i\rangle$, and |

| | |
|---|---|
| | $\left|\mathbf{s}'\right\rangle = F(\mathbf{T})\mathbf{s} = \sum_{n_{\min}} \lambda_i U_{2,i} F_{1,i}^{-1} \left|s_i\right\rangle.$ |
| $U_2^{-1}(\gamma_i)$ | The Fourier-transformed eigenchannel interference, $U_2^{-1}(\gamma_i) \in \mathcal{CN}\left(0, \mathbf{K}_{U_2^{-1}(\gamma_i)}\right)$, $\mathbf{K}_{U_2^{-1}(\gamma_i)} = \sigma_{\gamma_i}^2 = \mathbb{E}\left[\left|\gamma_i\right|^2\right]$, $\left|U_2^{-1}(\gamma_i)\right\rangle = U_2^{-1}\left(\sum_{j\neq i}^{n_{\min}} \lambda_j U_{2,j} F_{1,j}^{-1}\right)\left|s_j\right\rangle$. The variance $\sigma_\gamma^2 \to 0$, in the low-SNR regimes. |
| $z \in \mathcal{CN}\left(0, \sigma_z^2\right)$ | The variable of a single-carrier Gaussian CV state, $\left|\varphi_i\right\rangle \in \mathcal{S}$. Zero-mean, circular symmetric complex Gaussian random variable, $\sigma_z^2 = \mathbb{E}\left[\left|z\right|^2\right] = 2\sigma_{\omega_0}^2$, with i.i.d. zero mean, Gaussian random quadrature components $x, p \in \mathbb{N}\left(0, \sigma_{\omega_0}^2\right)$, where $\sigma_{\omega_0}^2$ is the variance. |
| $\Delta \in \mathcal{CN}\left(0, \sigma_\Delta^2\right)$ | The noise variable of the Gaussian channel $\mathcal{N}$, with i.i.d. zero-mean, Gaussian random noise components on the position and momentum quadratures $\Delta_x, \Delta_p \in \mathbb{N}\left(0, \sigma_\mathcal{N}^2\right)$, $\sigma_\Delta^2 = \mathbb{E}\left[\left|\Delta\right|^2\right] = 2\sigma_\mathcal{N}^2$. |
| $d \in \mathcal{CN}\left(0, \sigma_d^2\right)$ | The variable of a Gaussian subcarrier CV state, $\left|\phi_i\right\rangle \in \mathcal{S}$. Zero-mean, circular symmetric Gaussian random variable, $\sigma_d^2 = \mathbb{E}\left[\left|d\right|^2\right] = 2\sigma_\omega^2$, with i.i.d. zero mean, Gaussian random quadrature components $x_d, p_d \in \mathbb{N}\left(0, \sigma_\omega^2\right)$, where $\sigma_\omega^2$ is the modulation variance of the Gaussian subcarrier CV state. |
| $F^{-1}(\cdot) = \text{CVQFT}^\dagger(\cdot)$ | The inverse CVQFT transformation, applied by the encoder, continuous-variable unitary operation. |
| $F(\cdot) = \text{CVQFT}(\cdot)$ | The CVQFT transformation, applied by the decoder, continuous-variable unitary operation. |
| $F^{-1}(\cdot) = \text{IFFT}(\cdot)$ | Inverse FFT transform, applied by the encoder. |
| $\sigma_{\omega_0}^2$ | Single-carrier modulation variance. |
| $\sigma_\omega^2 = \frac{1}{l}\sum_l \sigma_{\omega_i}^2$ | Multicarrier modulation variance. Average modulation variance of the $l$ Gaussian sub-channels $\mathcal{N}_i$. |
| $\lvert\phi_i\rangle = \lvert\text{IFFT}(z_{k,i})\rangle$ $= \lvert F^{-1}(z_{k,i})\rangle = \lvert d_i\rangle.$ | The $i$-th Gaussian subcarrier CV of user $U_k$, where IFFT stands for the Inverse Fast Fourier Transform, $\lvert\phi_i\rangle \in \mathcal{S}$, $d_i \in \mathcal{CN}(0,\sigma_{d_i}^2)$, $\sigma_{d_i}^2 = \mathbb{E}[\lvert d_i\rvert^2]$, $d_i = x_{d_i} + \mathrm{i}p_{d_i}$, $x_{d_i} \in \mathbb{N}(0,\sigma_{\omega_F}^2)$, $p_{d_i} \in \mathbb{N}(0,\sigma_{\omega_F}^2)$ are i.i.d. zero-mean Gaussian random quadrature components, and $\sigma_{\omega_F}^2$ is the variance of the Fourier transformed Gaussian state. |
| $\lvert\varphi_{k,i}\rangle = \text{CVQFT}(\lvert\phi_i\rangle)$ | The decoded single-carrier CV of user $U_k$ from the subcarrier CV, expressed as $F(\lvert d_i\rangle) = \lvert F(F^{-1}(z_{k,i}))\rangle = \lvert z_{k,i}\rangle$. |
| $\mathcal{N}$ | Gaussian quantum channel. |
| $\mathcal{N}_i, i = 1,\ldots,n$ | Gaussian sub-channels. |
| $T(\mathcal{N})$ | Channel transmittance, normalized complex random variable, $T(\mathcal{N}) = \operatorname{Re} T(\mathcal{N}) + \mathrm{i}\operatorname{Im} T(\mathcal{N}) \in \mathcal{C}$. The real part identifies the position quadrature transmission, the imaginary part identifies the transmittance of the position quadrature. |
| $T_i(\mathcal{N}_i)$ | Transmittance coefficient of Gaussian sub-channel $\mathcal{N}_i$, $T_i(\mathcal{N}_i) = \operatorname{Re}(T_i(\mathcal{N}_i)) + \mathrm{i}\operatorname{Im}(T_i(\mathcal{N}_i)) \in \mathcal{C}$, quantifies the position and momentum quadrature transmission, with (normalized) real and imaginary parts $0 \leq \operatorname{Re} T_i(\mathcal{N}_i) \leq 1/\sqrt{2}$, $0 \leq \operatorname{Im} T_i(\mathcal{N}_i) \leq 1/\sqrt{2}$, where $\operatorname{Re} T_i(\mathcal{N}_i) = \operatorname{Im} T_i(\mathcal{N}_i)$. |
| $T_{Eve}$ | Eve's transmittance, $T_{Eve} = 1 - T(\mathcal{N})$. |
| $T_{Eve,i}$ | Eve's transmittance for the $i$-th subcarrier CV. |
| $\mathcal{A} \subseteq K$ | The subset of allocated users, $\mathcal{A} \subseteq K$. Only the allocated users can transmit information in a given (particularly the $j$-th) AMQD block. The cardinality of subset $\mathcal{A}$ is $\lvert\mathcal{A}\rvert$. |
| $U_k,\ k = 1,\ldots,\lvert\mathcal{A}\rvert$ | An allocated user from subset $\mathcal{A} \subseteq K$. |
| $\mathbf{z} = \mathbf{x} + \mathrm{i}\mathbf{p} = (z_1,\ldots,z_d)^T$ | A $d$-dimensional, zero-mean, circular symmetric complex random Gaussian vector that models $d$ Gaussian CV input states, $\mathcal{CN}(0,\mathbf{K_z})$, $\mathbf{K_z} = \mathbb{E}[\mathbf{zz}^\dagger]$, where $z_i = x_i + \mathrm{i}p_i$, |
| | $\mathbf{x} = (x_1,...,x_d)^T$, $\mathbf{p} = (p_1,...,p_d)^T$, with $x_i \in \mathbb{N}(0, \sigma^2_{\omega_0})$, $p_i \in \mathbb{N}(0, \sigma^2_{\omega_0})$ i.i.d. zero-mean Gaussian random variables. |
| $\mathbf{d} = F^{-1}(\mathbf{z})$ | An $l$-dimensional, zero-mean, circular symmetric complex random Gaussian vector of the $l$ Gaussian subcarrier CVs, $\mathcal{CN}(0, \mathbf{K_d})$, $\mathbf{K_d} = \mathbb{E}[\mathbf{dd}^\dagger]$, $\mathbf{d} = (d_1,...,d_l)^T$, $d_i = x_i + \mathrm{i}p_i$, $x_i, p_i \in \mathbb{N}(0, \sigma^2_{\omega_F})$ are i.i.d. zero-mean Gaussian random variables, $\sigma^2_{\omega_F} = 1/\sigma^2_{\omega_0}$. The $i$-th component is $d_i \in \mathcal{CN}(0, \sigma^2_{d_i})$, $\sigma^2_{d_i} = \mathbb{E}[|d_i|^2]$. |
| $\mathbf{y}_k \in \mathcal{CN}(0, \mathbb{E}[\mathbf{y}_k \mathbf{y}_k^\dagger])$ | A $d$-dimensional zero-mean, circular symmetric complex Gaussian random vector. |
| $y_{k,m}$ | The $m$-th element of the $k$-th user's vector $\mathbf{y}_k$, expressed as $y_{k,m} = \sum_l F(T_i(\mathcal{N}_i))F(d_i) + F(\Delta_i)$. |
| $F(\mathbf{T}(\mathcal{N}))$ | Fourier transform of $\mathbf{T}(\mathcal{N}) = [T_1(\mathcal{N}_1)...,T_l(\mathcal{N}_l)]^T \in \mathcal{C}^l$, the complex transmittance vector. |
| $F(\Delta)$ | Complex vector, expressed as $F(\Delta) = e^{\frac{-F(\Delta)^T \mathbf{K}_{F(\Delta)} F(\Delta)}{2}}$, with covariance matrix $\mathbf{K}_{F(\Delta)} = \mathbb{E}[F(\Delta)F(\Delta)^\dagger]$. |
| $\mathbf{y}[j]$ | AMQD block, $\mathbf{y}[j] = F(\mathbf{T}(\mathcal{N}))F(\mathbf{d})[j] + F(\Delta)[j]$. |
| $\tau = \|F(\mathbf{d})[j]\|^2$ | An exponentially distributed variable, with density $f(\tau) = (1/2\sigma^{2n}_\omega)e^{-\tau/2\sigma^2_\omega}$, $\mathbb{E}[\tau] \leq n 2\sigma^2_\omega$. |
| $T_{Eve,i}$ | Eve's transmittance on the Gaussian sub-channel $\mathcal{N}_i$, $T_{Eve,i} = \operatorname{Re} T_{Eve,i} + \mathrm{i}\operatorname{Im} T_{Eve,i} \in \mathcal{C}$, $0 \leq \operatorname{Re} T_{Eve,i} \leq 1/\sqrt{2}$, $0 \leq \operatorname{Im} T_{Eve,i} \leq 1/\sqrt{2}$, $0 \leq |T_{Eve,i}|^2 < 1$. |
| $R_k$ | Transmission rate of user $U_k$. |
| $d_i$ | A $d_i$ subcarrier in an AMQD block. For subset $\mathcal{A} \subseteq K$ with $|\mathcal{A}|$ users and $l$ Gaussian sub-channels for the transmission, $d_i = \frac{1}{\sqrt{n}} \sum_{k=1}^{|\mathcal{A}|} z_k e^{\frac{-\mathrm{i}2\pi ik}{n}}, i = 1,...,l$. |

| | |
|---|---|
| $\nu_{\min}$ | The $\min\{\nu_1,...,\nu_l\}$ minimum of the $\nu_i$ sub-channel coefficients, where $\nu_i = \sigma_{\mathcal{N}}^2 \big/ \left|F\left(T_i\left(\mathcal{N}_i\right)\right)\right|^2$ and $\nu_i < \nu_{Eve}$. |
| $\sigma_\omega^2$ | Modulation variance, $\sigma_\omega^2 = \nu_{Eve} - \nu_{\min}\mathcal{G}(\delta)_{p(x)}$, where $\nu_{Eve} = \frac{1}{\lambda}$, $\lambda = \left|F\left(T_{\mathcal{N}}^*\right)\right|^2 = \frac{1}{n}\sum_{i=1}^{n}\left|\sum_{k=1}^{n}T_k^* e^{\frac{-i2\pi ik}{n}}\right|^2$ and $T_{\mathcal{N}}^*$ is the expected transmittance of the Gaussian sub-channels under an optimal Gaussian collective attack. |
| $\nu_\kappa$ | Additional sub-channel coefficient for the correction of modulation imperfections. For an ideal Gaussian modulation, $\nu_\kappa = 0$, while for an arbitrary $p(x)$ distribution $\nu_\kappa = \nu_{\min}\left(1 - \mathcal{G}(\delta)_{p(x)}\right)$, where $\kappa = \frac{1}{\nu_{Eve} - \nu_{\min}\left(\mathcal{G}(\delta)_{p(x)} - 1\right)}$. |
| $\mathcal{N}_{U_k}[j] = [\mathcal{N}_1,...,\mathcal{N}_s]^T$ | The set of $\mathcal{N}_i$ Gaussian sub-channels from the set of $l$ good sub-channels that transmit the $s$ subcarriers of user $U_k$ in the $j$-th AMQD block. |
| $\sigma_{\omega_i'}^2$ | The constant modulation variance $\sigma_{\omega_i'}^2$ for eigenchannel $\lambda_i$, evaluated as $\sigma_{\omega_i'}^2 = \mu - \left(\sigma_{\mathcal{N}}^2 \big/ \max_{n_{\min}}\lambda_i^2\right) = \frac{1}{n_{\min}}\sigma_{\omega'}^2$, with a total constraint $\sigma_{\omega'}^2 = \sum_{n_{\min}}\sigma_{\omega_i'}^2 = \frac{1}{l}\sum_l \sigma_{\omega_i}^2 = \sigma_\omega^2$. |
| $\sigma_{\omega''}^2$ | The modulation variance of the AMQD multicarrier transmission in the SVD environment. Expressed as $\sigma_{\omega''}^2 = \nu_{Eve} - \left(\sigma_{\mathcal{N}}^2 \big/ \max_{n_{\min}}\lambda_i^2\right)$, where $\lambda_i$ is the $i$-th eigenchannel of $F(\mathbf{T})$, $\max_{n_{\min}}\lambda_i^2$ is the largest eigenvalue of $F(\mathbf{T})F(\mathbf{T})^\dagger$, with a total constraint $\frac{1}{l}\sum_l \sigma_{\omega_i''}^2 = \sigma_{\omega''}^2 > \sigma_\omega^2$. |
| $\mathbf{K}_{k,\mathbf{s}}$ | The covariance matrix $\mathbf{K}_{k,\mathbf{s}}$ of user $U_k$. In the first setting of AMQD-MQA it is expressed as $\mathbf{K}_{k,\mathbf{s}} = F_1 diag\left\{\sigma_{\omega_{k,1}'}^2,...,\sigma_{\omega_{k,n_{\min}}'}^2\right\}F_1^{-1}$. In the second setting, it is evaluated as $\mathbf{K}_{k,\mathbf{s}} = U_1 diag\left\{\sigma_{\omega_{k,1}'}^2,...,\sigma_{\omega_{k,n_{\min}}'}^2\right\}U_1^{-1}$. |
| $\nu'_{\min}$ | The minimum of the $\nu'_i$-s of the $l$ Gaussian sub-channels $\mathcal{N}_i$, obtained at the SVD of $F(\mathbf{T})$, $\nu'_{\min} = \frac{\sigma_\mathcal{N}^2}{\max_{n_{\min}} \lambda_i^2}$. |
| $\Pi$ | Difference of $\nu_{\min}$ and $\nu'_{\min}$, $\Pi = \nu_{\min} - \nu'_{\min} = \sigma_\mathcal{N}^2 \Big/ \max_{\forall i}\big|F(T_i(\mathcal{N}_i))\big|^2 - \sigma_\mathcal{N}^2 \Big/ \max_{n_{\min}} \lambda_i^2$. |
| $S(F(\mathbf{T}))$ | The statistical model of $F(\mathbf{T})$ at a partial channel side information, $S(F(\mathbf{T})) = \xi_{K_{out}}^{-1} \Gamma \xi_{K_{in}}$, where $\xi_{K_{out}}^{-1}$ and $\xi_{K_{in}}$ are unitaries that formulate the input covariance matrix $\mathbf{K_s} = \xi_{K_{in}} \wp \xi_{K_{in}}^{-1}$, while $\wp$ is a diagonal matrix, $\mathbf{K_s} = Q diag\{\sigma_{\omega'_1}^2, \ldots, \sigma_{\omega'_{K_{in}}}^2\} Q^\dagger$. |
| $\wp = \sigma_{\omega'}^2 \frac{1}{K_{in}} \mathbf{I}_{K_{in}}$ | Diagonal matrix, the entries of $\wp$ are chosen such that $\wp = \sigma_{\omega'}^2 \frac{1}{K_{in}} \mathbf{I}_{K_{in}}$, where $\sigma_{\omega'}^2 = \sum_{n_{\min}} \sigma_{\omega'_i}^2$ is the total constraint. |
| $\mathcal{P}_i$ | Projector, $\mathcal{P}_i(|s'_i\rangle) = \text{II}_i^\perp$, where set $\text{II}_i^\perp = \{|s'_j\rangle\}_{j=1}^{n_{\min}}$, $j \neq i$ consists of $n_{\min}$ CV states $|s'_j\rangle$, such that each $|s'_j\rangle$ is orthogonal to the sub-space $\text{II}_i$ spanned by the $n_{\min} - 1$ eigenchannels $\{\lambda_j U_{2,j} F_{1,j}^{-1}\}$, $j = 1, \ldots, n_{\min}, j \neq i$. |
| $W_i$ | The operator of the decoder-side interference cancellation, $W_i = \mathcal{P}_i^\dagger \mathcal{P}_i(\lambda_i U_{2,i} F_{1,i}^{-1})$. |
| $V$ | Encoder-side interference cancellation operator. Defines a phase space constellation $\mathcal{C}_\mathcal{S}$ and performs the precoding. |
| $|\kappa_i\rangle$ | Naive precoding with unbounded additional modulation variance, $|\kappa_i\rangle = |\varphi_i\rangle + (-U_2^{-1}|\gamma_i\rangle) = |\varphi_i\rangle - U_2^{-1}|\gamma_i\rangle$. |
| $|\kappa_i\rangle$ | Precoding with a boundary condition, $|\kappa_i\rangle = \min\{EC_k|\varphi_i\rangle - U_2^{-1}|\gamma_i\rangle\}$. |
| $|\tilde{\kappa}_i\rangle$ | Optimal precoding with a boundary condition and maximum likelihood nearest neighbor decision rule at $\mathcal{D}_\alpha$, $|\tilde{\kappa}_i\rangle = \min\{EC_k|\varphi_i\rangle - \alpha U_2^{-1}|\gamma_i\rangle\}$. |

| | |
|---|---|
| $EC_k\lvert\varphi_i\rangle$ | Equivalence class of $\lvert\varphi_i\rangle$. Identifies a replication of $\lvert\varphi_i\rangle$ in the phase space by $EC_k\lvert\varphi_i\rangle$ of $\Omega$. |
| $\alpha$ | Parameter of a maximum likelihood nearest neighbor decision decoder $\mathcal{D}_\alpha$, $\alpha = \sigma_\omega^2 \big/ \sigma_\omega^2 + \sigma_\mathcal{N}^2$. |
| $\Omega$ | Equivalence class set, $\Omega \equiv \{EC_k\lvert\varphi_i\rangle\}_{k=1}^{d_{EC}}$. |
| $\mathcal{C}_\mathcal{S}$ | Phase space constellation $\mathcal{C}_\mathcal{S}$ constructed by $V$. Superset of $\Omega$ equivalence classes $\mathcal{C}_\mathcal{S} = \bigcup_D \Omega = \bigcup_D \{EC_k\lvert\varphi_i\rangle\}_{k=1}^{d_{EC}}$. |
| $\varsigma = \{\lvert\varphi_i\rangle\}_{i=1}^d$ | A $d$-dimensional random phase space code. |
| $r$ | Radius of an uncertainty sphere around $\alpha\lvert\varphi_i'\rangle$, in which the transmit Gaussian $\lvert\varphi_i\rangle$ lies, $r = \sqrt{\frac{\sigma_\omega^2 \sigma_\mathcal{N}^2}{\sigma_\omega^2 + \sigma_\mathcal{N}^2}}$. |
| $\mathcal{D}_\alpha$ | Maximum likelihood nearest neighbor decision rule decoder, with $\alpha$ constraint, $\alpha = \frac{\sigma_\omega^2}{\sigma_\omega^2 + \sigma_\mathcal{N}^2}$, and error probability $p_{err} = \frac{\sigma_\omega^2 \sigma_\mathcal{N}^2}{\sigma_\omega^2 + \sigma_\mathcal{N}^2}$. |
| $h_{\min}$ | Minimal distance between arbitrary constellation points, $h_{\min} = 2\sigma_\omega$. |
| $\chi_i$ | The sum of terms $U(\Delta_i) \in \mathcal{CN}(0, \mathbf{K}_{\Delta_i})$ and $U_2^{-1}(\gamma_i) \in \mathcal{CN}(0, \mathbf{K}_{U_2^{-1}(\gamma_i)})$, $\chi_i = U_2^{-1}(\gamma_i) + U(\Delta_i)$, $\chi_i \in \mathcal{CN}(0, \mathbf{K}_{\chi_i})$, where $\mathbf{K}_{\chi_i} = \mathbf{K}_{\Delta_i} + \mathbf{K}_{U_2^{-1}(\gamma_i)} = \sigma_\mathcal{N}^2 \mathbf{I}_{K_{out}} + \sum_{j \neq i}^{K_{in}} \sigma_{\omega_j''}^2 F(\mathbf{T}_j(\mathcal{N}_{U_j})) F(\mathbf{T}_j(\mathcal{N}_{U_j}))^\dagger$, and $\mathcal{D}_\alpha = \mathbf{K}_{\chi_i}^{-1} F(\mathbf{T}_i(\mathcal{N}_{U_i}))$. |
| $Q$ | The Gaussian tail function, $Q\left(\sqrt{\frac{1}{2}\sum_l(\nu_{Eve} - \lvert\delta_i\rvert^2)}\right)$. |
| $\mathcal{C}_\mathcal{S}^P(\mathcal{N})$ | Random phase space permutation constellation for the transmission of the Gaussian subcarriers, expressed as $\mathcal{C}_\mathcal{S}^P(\mathcal{N}) = \mathcal{C}_\mathcal{S}^P(\mathcal{N}_1),...,\mathcal{C}_\mathcal{S}^P(\mathcal{N}_l) = \left(\Big\lvert\phi_{1...d_{\mathcal{C}_\mathcal{S}^P(\mathcal{N}_1)}}\Big\rangle, P_2\Big\lvert\phi_{1...d_{\mathcal{C}_\mathcal{S}^P(\mathcal{N}_2)}}\Big\rangle,...,P_l\Big\lvert\phi_{1...d_{\mathcal{C}_\mathcal{S}^P(\mathcal{N}_l)}}\Big\rangle\right)$, where $\lvert\phi_i\rangle$ are the Gaussian subcarrier CVs, $P_i$, $i=2,...,l$ |
| | is a random permutation operator, $d_{\mathcal{C}_\mathcal{S}(\mathcal{N}_i)} = d_{\mathcal{C}_\mathcal{S}(\mathcal{N}_j)}$ is the cardinality of $\mathcal{C}_\mathcal{S}(\mathcal{N}_i)$. The optimality function is $o(\mathcal{C}_\mathcal{S}(\mathcal{N})) = \sum_l \left(\nu_{Eve} - |\delta_i|^2\right)$. |
| $d_{\mathcal{C}_\mathcal{S}^P(\mathcal{N}_i)}$ | Cardinality of $\mathcal{C}_\mathcal{S}(\mathcal{N}_i)$. |
| $\delta_i$ | The normalized difference of two Gaussian subcarriers $d_{A,i}$ and $d_{B,i}$, $\delta_i = \frac{1}{\sqrt{\frac{\sigma_{\omega''}^2}{\sigma_\mathcal{N}^2}}}(d_{A,i} - d_{B,i})$. |
| $o(\mathcal{C}_\mathcal{S}(\mathcal{N}))$ | The optimality function, quantifies the optimality of the phase space constellation $\mathcal{C}_\mathcal{S}(\mathcal{N})$, as $o(\mathcal{C}_\mathcal{S}(\mathcal{N})) = \sum_l \left(\nu_{Eve} - |\delta_i|^2\right)$. |
| $\partial$ | Difference function of Gaussian subcarriers (phase space symbols) $d_{A,i}$ and $d_{A,j}$ in constellations $\mathcal{C}_\mathcal{S}^P(\mathcal{N}_k)$, $k = 1,...,l$. For two Gaussian sub-channels $\mathcal{N}_1$ and $\mathcal{N}_2$, $\partial(\mathcal{C}_\mathcal{S}^P(\mathcal{N}_1)) = \min_{\forall d_{A,i}}(d_{A,i} - d_{A,j})$, $\partial(\mathcal{C}_\mathcal{S}^P(\mathcal{N}_2)) = \mho \cdot \partial(\mathcal{C}_\mathcal{S}^P(\mathcal{N}_1))$, where $\mho \geq 2$, $j \neq i$. |
| $\tilde{z}_i$ | Estimation of input $z_i$ at decoder $\mathcal{D}_\alpha$, such that $\tilde{z}_i = A\mathbf{c}^\dagger \mathbf{y}_i = A\mathbf{c}^\dagger F(\mathbf{T}_i(\mathcal{N}_{U_i}))z_i + A\mathbf{c}^\dagger \chi_i$, minimizes the mean square error $\mathbb{E}\left[|z_i - \tilde{z}_i|^2\right]$, where $A = \frac{\mathbb{E}[|z_i|^2]\left|\mathbf{c}^\dagger F(\mathbf{T}_i(\mathcal{N}_{U_i}))\right|^2}{\mathbb{E}[|z_i|^2]\left|\mathbf{c}^\dagger F(\mathbf{T}_i(\mathcal{N}_{U_i}))\right|^2 + \mathbf{c}^\dagger \mathbf{K}_{\chi_i} \mathbf{c}} + \frac{F(\mathbf{T}_i(\mathcal{N}_{U_i}))^\dagger \mathbf{c}}{\left|F(\mathbf{T}_i(\mathcal{N}_{U_i}))^\dagger \mathbf{c}\right|}$, $\mathbf{c} = \mathbf{K}_{\chi_i}^{-1} F(\mathbf{T}_i(\mathcal{N}_{U_i}))$, and $\mathbf{K}_{\chi_i} = \sigma_\mathcal{N}^2 \mathbf{I}_{K_{out}}$ in the low-SNR regimes. |

## S.2 Abbreviations

| | |
|---|---|
| **AMQD** | Adaptive Multicarrier Quadrature Division |
| **AWGN** | Additive White Gaussian Noise |
| **CV** | Continuous-Variable |
| **CVQFT** | Continuous-Variable Quantum Fourier Transform |
| **DV** | Discrete-Variable |
| **EC** | Equivalence Class |
| **FFT** | Fast Fourier Transform |
| **IFFT** | Inverse Fast Fourier Transform |
| **MQA** | Multiuser Quadrature Allocation |
| **OFDM** | Orthogonal Frequency-Division Multiplexing |
| **OFDMA** | OFDM Multiple Access |
| **QKD** | Quantum Key Distribution |
| **SIA** | Singular Interference Avoider |
| **SNR** | Signal-to-Noise Ratio |
| **SNIR** | Signal-to-Noise-plus-Interference Ratio |
| **SVD** | Singular Value Decomposition |